\documentclass[sigconf]{acmart}

\setcopyright{none}
\copyrightyear{2027}
\acmYear{2027}
\acmConference[CHI '27]{CHI Conference on Human Factors in Computing
  Systems}{April 2027}{TBD}
\acmBooktitle{CHI Conference on Human Factors in Computing Systems
  (CHI '27), April 2027}
\acmDOI{XX.XXXX/XXXXXXX.XXXXXXX}
\acmISBN{978-X-XXXX-XXXX-X/27/04}

\usepackage{booktabs}
\usepackage{graphicx}

\begin{document}

\title{TouchTwin: Human--AI Haptic Authoring for 3D-Scanned Tabletop
  Scenes through Language and Touch}

\author{Wanli Qian}
\affiliation{%
  \institution{University of Southern California}
  \city{Los Angeles}
  \country{USA}}
\email{wanliqia@usc.edu}

\author{Xinlei Yu}
\affiliation{%
  \institution{University of Southern California}
  \city{Los Angeles}
  \country{USA}}
\email{xinleiyu@usc.edu}

\author{Heather Culbertson}
\affiliation{%
  \institution{University of Southern California}
  \city{Los Angeles}
  \country{USA}}
\email{hculbert@usc.edu}

\renewcommand{\shortauthors}{Anon. et al.}

\begin{abstract}
Creating haptic experiences from real-world environments remains difficult. Visual scans capture appearance but not material feel, while manual haptic
authoring requires specialized expertise. We present TouchTwin, a human--AI
approach that turns phone scans of tabletop scenes into editable haptic
scenes through a propose–feel–revise workflow. AI proposes named material regions that users experience through touch and revise using language and spatial edits; a model grounded in measured tool–surface interactions converts material descriptions into haptic responses. In a study with 12 participants and physical
reference objects present, automatic drafts scored $5.4/7$ as useful starting points despite only $4.0/7$ realism. After a median of 6.6 minutes of
editing, realism increased to $5.6/7$, and material-family accuracy from 52\% to 92\%. Participants also used the same workflow to author material
feels without a physical reference. Our findings show how imperfect AI predictions can support haptic authoring when users can experience and directly revise their perceptual consequences.
\end{abstract}

\ccsdesc[500]{Human-centered computing~Haptic devices}
\ccsdesc[300]{Human-centered computing~Interaction techniques}
\ccsdesc[500]{Computing methodologies~Computer vision}

\keywords{haptics, texture rendering, authoring tools, vision--language
models, 3D reconstruction, digital twins, telepresence,
human-in-the-loop, human--AI collaboration, generative AI}

\begin{teaserfigure}
  \centering
  \includegraphics[width=0.98\textwidth]{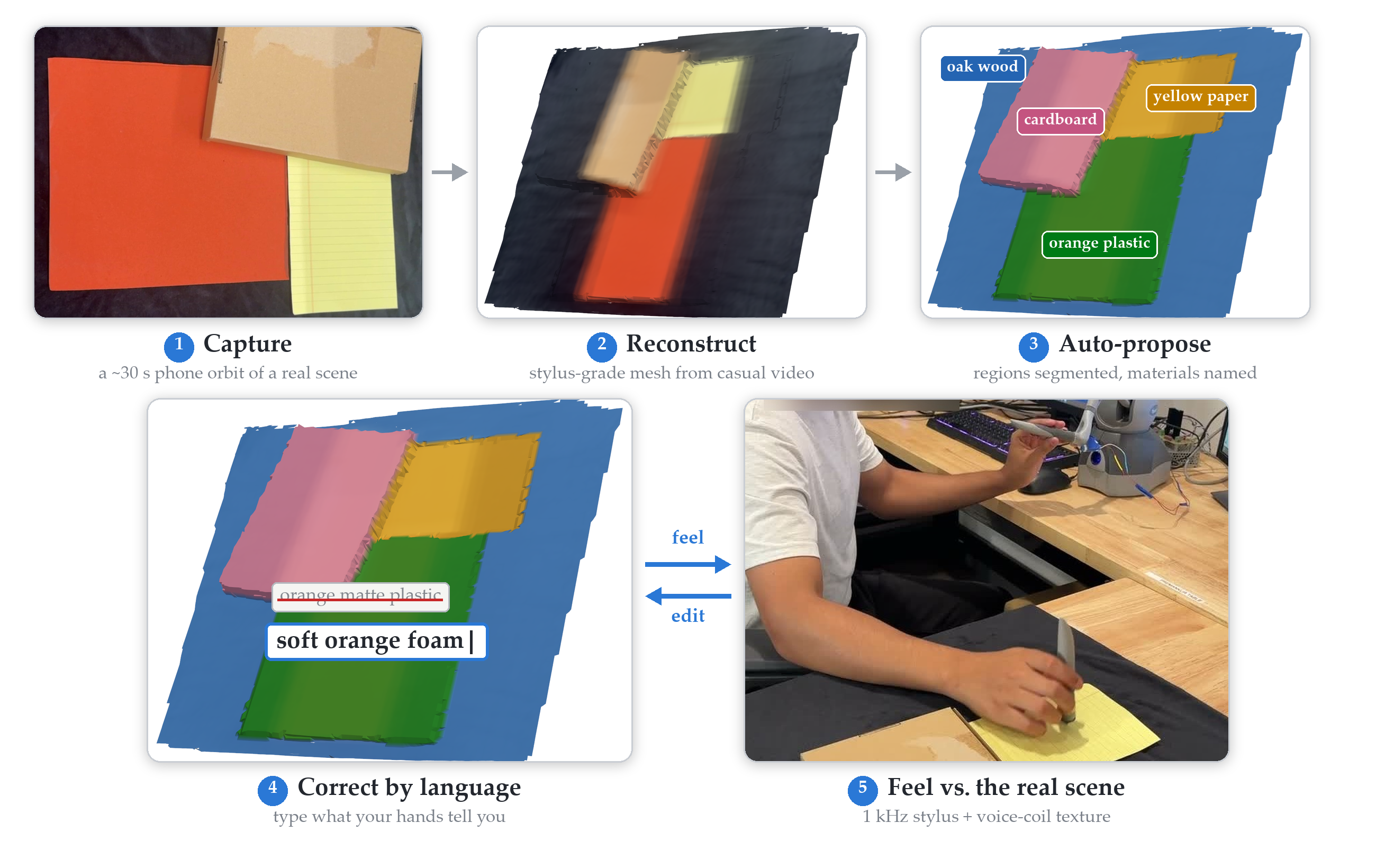}
  \caption{TouchTwin turns a phone video of a tabletop scene into a
  touchable twin: a stack of editable haptic texture layers (shown: the
  study's harder scene---a cardboard box, a lined yellow pad, and an
  orange foam sheet on dark velvet). The system drafts regions and
  material names automatically (3); the velvet reads as ``oak wood''
  and the foam as ``orange plastic''---compliance is invisible to a
  vision model. The user revises the draft by typing material names and
  repainting boundaries (4, the study's real correction), then feels
  the result on a force-feedback stylus with a voice-coil texture
  channel, with the physical objects on the desk for comparison (5).}
  \Description{Five-panel pipeline overview. First, a frame from the
  phone capture video showing an orange foam sheet, a cardboard box,
  and a lined yellow pad on dark velvet cloth. Second, the
  reconstructed 3D mesh of the same scene. Third, the mesh with four
  colored material regions labeled oak wood, cardboard, yellow paper,
  and orange plastic. Fourth, the same scene with the label orange
  matte plastic struck through and the text soft orange foam being
  typed as a replacement. Fifth, a photograph of a participant, face
  blurred, holding the force-feedback stylus with the same physical
  objects on the desk.}
  \label{fig:teaser}
\end{teaserfigure}

\maketitle

\section{Introduction}

A phone video can capture how a place looks, but not how it feels. Modern
reconstruction methods~\cite{mildenhall2020nerf,kerbl2023gaussian,wang2025vggt}
can turn casual scans into detailed 3D environments for virtual tours,
e-commerce, and telepresence~\cite{orts2016holoportation}, yet recreating how those
environments feel remains a largely manual authoring problem. Each surface
must be associated with haptic properties that a device can render during
interaction, requiring expertise that most users do not have. Automating
this process is appealing, but errors in inferred material properties can
become immediately apparent through touch: a cardboard box that looks
plausible but feels like wood is still a poor haptic representation. This
creates an interaction-design challenge beyond simply improving recognition
accuracy: how can imperfect automation help people create convincing haptic
experiences while still giving them intuitive ways to identify and correct
what the system gets wrong?

Existing approaches occupy two ends of this automation--control spectrum.
Vision--language models (VLMs) offer a way to infer material properties
directly from visual scans, but their predictions remain unreliable in
unconstrained scenes. On a small six-scene benchmark, for example, a locally
runnable VLM (Qwen2.5-VL-3B~\cite{bai2025qwen25vl}) identified the exact
material for 13 of 24 surfaces. At the other extreme, manual haptic authoring
tools give users precise control, but typically require them to specify or
tune low-level properties such as friction, stiffness, and vibration
spectra~\cite{schneider2017haptic,ryu2008posvibeditor,schneider2016studying,pezent2021syntacts}.
Neither approach is well suited to non-experts who want to reproduce
how a scanned environment feels: full automation provides too little control
when predictions fail, while manual authoring demands too much haptic
expertise.

Rather than treating material recognition as a final prediction, we treat
it as the start of an interaction. TouchTwin uses automation to create an
editable haptic draft that users can feel, evaluate, and revise. The system
proposes material regions and their haptic responses, while users correct
errors by renaming materials or repainting region boundaries. Crucially,
these corrections operate at the same semantic level as the system's
predictions: users can touch a box that feels wrong and type ``cardboard,''
rather than manipulating friction coefficients or vibration spectra. This
propose--feel--revise workflow combines automatic scene interpretation with
human perceptual judgment and control, allowing imperfect recognition to
remain useful without requiring users to become haptic experts.

TouchTwin implements this approach by turning a casual phone scan into an
editable haptic scene. The system proposes surface regions and material
descriptions, then uses a model trained on measured material interactions to
generate corresponding haptic responses. Users can feel the scene and revise
the descriptions or region boundaries, with each edit updating the rendered
haptics. TouchTwin targets material-category fidelity: it aims to make each
region feel like the right kind of material, rather than reproduce the exact
vibration and tapping response of the scanned physical specimen.

We evaluated this workflow with 12 participants who compared the generated
scenes directly with the physical objects. Although the automatic proposals
were only partially accurate, participants rated them $5.4/7$ as useful
starting points. After a median of 6.6 minutes of editing, perceived realism
increased from $4.0$ to $5.6/7$, and scoring the saved material layers
against ground truth showed material-family accuracy increasing from 52\% to
92\%.
Generated-model distance to the individual reference samples, however, did
not reliably decrease.

We further explored how the same interaction supports both reproducing
existing scenes and creating new haptic experiences. Participants used
freeform material descriptions to author textures without a physical
referent, while generally preferring this mode for creative work and the
automatic proposals for reproducing a real scene. Together, these results
suggest a broader role for AI in haptic authoring: rather than needing to
produce a correct final experience, AI can provide semantically meaningful
starting points that people can experience, interpret, and refine through
touch.

Our contributions are (1) a propose--feel--revise interaction model for
human--AI haptic authoring, (2) TouchTwin, an end-to-end system realizing
this approach for 3D-scanned tabletop scenes, and (3) empirical evidence
that users can improve imperfect automatic haptic drafts through brief
semantic and spatial edits.

\section{Related Work}

\subsection{From Haptic Texture Rendering to Touchable 3D Scenes}

Rendering convincing virtual textures requires reproducing perceptual cues
that arise during physical contact. Fine texture is largely perceived
through vibrations generated during surface exploration. The duplex theory
distinguishes coarse spatial cues from a vibration-sensitive channel that
depends on motion~\cite{hollins2000evidence}, and scanning vibrations and
their Pacinian-weighted spectra help explain differences among fine
textures~\cite{bensmaia2003vibrations,bensmaia2005pacinian}. These cues
remain important during tool-mediated interaction: a rigid probe
preserves texture perception~\cite{klatzky2003feeling,yoshioka2007texture}
along dimensions such as roughness, hardness, and
stickiness~\cite{okamoto2013psychophysical,bergmanntiest2010tactual,lederman2009haptic},
and users tend to attribute stylus vibrations to the contacted
surface~\cite{choi2013vibrotactile}.

Data-driven rendering exploits these effects by recording physical
tool--surface interactions and resynthesizing them during virtual
contact~\cite{romano2012creating}. The Penn Haptic Texture Toolkit, for
example, models 100 measured materials using force- and speed-dependent
autoregressive (AR) vibration
models~\cite{culbertson2014modeling,culbertson2014hundred}. Subsequent work
incorporates contact transients and speed-dependent vibration to improve
realism~\cite{kuchenbecker2006improving,culbertson2015should}, and explores
alternative learned
representations~\cite{abdulali2016datadriven,joolee2022data,heravi2024development}.
Other displays reproduce texture cues through electrovibration, voice coils,
or audio-driven
actuators~\cite{bau2010teslatouch,strohmeier2017generating,pezent2021syntacts}.
These methods generally assume that a material model has already been
measured or selected, often through instrumented
capture~\cite{pai2001scanning,kuchenbecker2011haptography,burka2016proton,otaduy2004perceptual}.
Together, these approaches provide increasingly rich ways to render how a
virtual surface feels once its haptic properties have been specified.

Extending these techniques from individual surfaces to complete virtual
environments introduces a different problem: how should haptic properties
be assigned across a 3D scene? Prior systems make scenes touchable through
passive or retargeted physical
proxies~\cite{simeone2015substitutional,hettiarachchi2016annexing,azmandian2016haptic,cheng2017sparse},
encountered-type devices~\cite{araujo2016snake,suzuki2020roomshift}, shape
displays~\cite{leithinger2014physical}, or fabricated
microgeometry~\cite{degraen2019enhancing,degraen2021capturing}, as well as
feedback rendered by grounded force-feedback arms~\cite{massie1994phantom}
and handheld haptic
controllers~\cite{whitmire2018haptic,choi2018claw,lee2019torc}, among
other approaches~\cite{culbertson2018haptics,weiss2023using}.
Some approaches explicitly connect captured physical environments to
virtual ones; Oasis, for example, uses matched real objects to make
reconstructed rooms physically explorable~\cite{sra2016procedurally}.
These approaches generally depend on matching physical matter, explicitly
authored haptic properties, or material models that have already been
measured or selected.

TouchTwin instead addresses the upstream problem of assigning haptic
properties to a casually captured scene that has not been instrumented for
haptic measurement. It associates scene regions with material descriptions
and maps those descriptions to models grounded in previously measured
tool--surface interactions. This trades specimen-level fidelity for easier
capture: a phone scan can provide the starting point for a touchable scene
without measuring every surface individually. The remaining challenge is
one of authoring: how people can specify, evaluate, and revise material feel
throughout the scene without manipulating haptic signals directly.

\subsection{Authoring Haptics: From Signal Timelines to Language}

Haptic authoring remains difficult for non-experts because designers must
translate intended experiences into device- and signal-level
parameters~\cite{schneider2017haptic,seifi2019haptipedia}. Prior tools move
beyond direct signal editing through example-driven
editors~\cite{ryu2008posvibeditor,schneider2016studying}, tactile
animation~\cite{schneider2015tactile}, crowdsourced proxy
ratings~\cite{schneider2016hapturk}, mid-air haptic design
tools~\cite{seifi2023feellustrator}, adaptive
tactons~\cite{john2024adaptics}, and vocalized
prototyping~\cite{degraen2021weirding}.
Language is particularly promising because people describe haptic
experiences using sensory and metaphorical terms rather than signal
parameters~\cite{seifi2015vibviz}. A taxonomy of 79 haptic authoring tools
documents natural-language description as a growing interaction approach
and identifies broader gaps in multimodal authoring
support~\cite{regimbal2026understanding}. Therefore, we designed TouchTwin to use a named
material attached to a spatial region as its primary authoring unit, rather
than exposing waveforms or spectra.

Recent generative approaches make semantic authoring increasingly feasible
by mapping higher-level inputs to haptic signals. Prior systems include
GAN-generated pen
vibrations~\cite{ujitoko2018vibrotactile,zhang2025texsensegan}, HapticGen's
text-conditioned vibration~\cite{sung2025hapticgen}, ChatHAP's conversational
generation and modification of vibrotactile
effects~\cite{lim2025chathap}, latent-diffusion
approaches~\cite{xiong2026hapticldm}, and perceptually trained
audio-to-vibration mappings~\cite{li2026sound2hap}. Related work generates
visual--tactile patches, printable microgeometry, 3D forms, and shape-display
behaviors~\cite{gao2023controllable,gao2024tactile,faruqi2025tactstyle,qian2024shapeit}.
These systems primarily generate individual effects or artifacts rather
than assigning and revising material feel across a captured 3D scene.

TouchTwin builds on Qian et al.'s language-guided texture
model~\cite{qian2026language}, which maps text descriptions to force- and
speed-dependent vibration and tapping responses through a latent space
trained on measured tool--surface interactions. Their work demonstrated
language-based generation for isolated material patches on a kinesthetic
haptic device, but did not address captured scenes, spatial segmentation,
or correction of automatically inferred materials. Scene2Hap is closer in
scene-level scope: it assigns vibrotactile feedback throughout authored VR
environments and models propagation between
objects~\cite{jingu2026scene2hap}, but does not provide a human correction
loop or ground its feedback in measured contact interactions.

TouchTwin uses language not only as an input for human authoring, but also
as a shared representation between human and AI. Automatic descriptions
and user revisions pass through the same text-to-haptics path. Combined
with editable spatial regions, this shared representation lets the same
interface support both automatic proposals and human revisions. TouchTwin
therefore uses language to support an iterative human--AI process in which
generated haptics can be felt, interpreted, and revised.

\subsection{Material Recognition as an Editable Proposal}

Automatically assigning haptic properties to a captured scene requires
inferring material information from visual input. Material recognition has
been supported by large-scale image datasets~\cite{bell2015material},
contrastive vision--language models~\cite{radford2021learning}, promptable
segmentation~\cite{kirillov2023segment,ravi2024sam2}, and instruction-tuned
VLMs~\cite{bai2025qwen25vl}. Other work connects material appearance with
contact acceleration and sound~\cite{strese2017multimodal}, with tactile
measurements~\cite{yuan2017connecting,gao2022objectfolder,yang2022touch},
and with language~\cite{yu2024octopi},
while multi-view reconstruction and depth completion recover 3D scenes from
casual phone video~\cite{wang2025vggt,murre2025}. These methods provide the
ingredients for proposing what surfaces might be made of. However, visual
recognition cannot observe every property relevant to touch, such as
compliance, and plausible visual predictions can produce conspicuous haptic
errors. TouchTwin therefore treats recognition as an editable proposal.

This approach builds on mixed-initiative human--AI interaction, in which
automation assists users while leaving its outputs understandable and
overridable~\cite{horvitz1999principles,amershi2019guidelines}. Simply adding
human correction does not guarantee better team
performance~\cite{bansal2021does}. TouchTwin presents a setting in which
human and AI have complementary information. The vision model proposes
materials across a scene, while a person with access to the physical
referent can compare those generated haptics with the real surfaces. The
user then corrects the same semantic unit that the AI proposed---a named material
region---through language or spatial editing.

Related 3D authoring systems also combine generated or inferred content
with human control. Language instructions revise
NeRFs~\cite{haque2023instructnerf2nerf}, construct mixed-reality
environments~\cite{delatorre2024llmr}, edit radiance
fields~\cite{vachha2025dreamcrafter}, and support generative 3D
design~\cite{liu20233dalle}. Style2Fab provides a relevant interaction
precedent; it automatically segments meshes and classifies segments as
functional or aesthetic, after which users can review and revise those
classifications before applying generative
styles~\cite{faruqi2023style2fab}. TouchTwin extends this
proposal--and--correction approach to haptic scene authoring, where users can
revise both the material description and the region to which it applies.

Taken together, prior work provides many of the capabilities needed for
AI-assisted haptic scene authoring, but not the interaction loop we study.
Haptic rendering can resynthesize measured material responses,
language-based generation makes those responses accessible without
signal-level expertise, and visual models propose material properties from
captured scenes. TouchTwin connects these capabilities through named
material regions that AI can propose, people can evaluate through touch,
and people can revise through language and spatial editing. We found no
end-to-end system suitable as a direct baseline; our comparisons therefore
use the automatic proposal before revision and the naming baselines in
\S\ref{sec:techeval}. A rated oracle-label condition would be needed to
separate the effects of recognition error from the remaining rendering gap
(\S\ref{sec:discussion}).

\section{TouchTwin: A Propose--Feel--Revise Haptic Authoring System}
\label{sec:system}

\begin{figure*}
  \centering
  \includegraphics[width=0.98\textwidth]{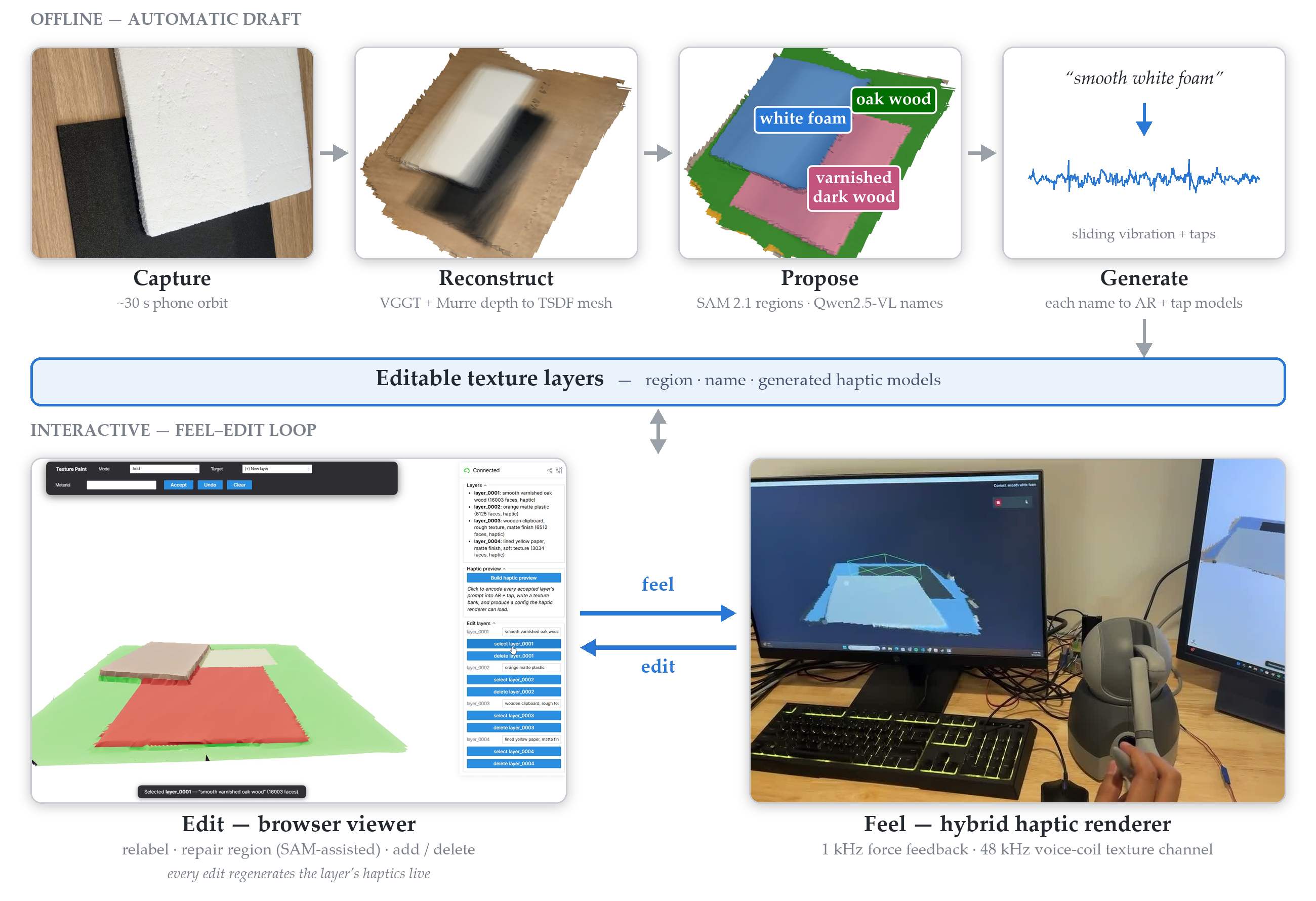}
  \caption{System pipeline. Offline (top): feed-forward reconstruction
  with diffusion depth completion, SAM~2.1 region lifting, VLM material
  naming, and language-conditioned generation of haptic texture
  models~\cite{qian2026language} produce a stack of editable texture
  layers. Interactive (bottom): the browser viewer's feel--edit
  correction loop and the hybrid haptic renderer (1\,kHz force-feedback
  servo plus voice-coil texture channel); every edit regenerates the
  layer's haptic models live.}
  \Description{Photo-based pipeline diagram. An offline row shows four
  stages with real thumbnails: a phone capture frame of a foam block
  and dark board; the reconstructed mesh of the same scene; the mesh
  with colored regions labeled white foam, oak wood, and varnished
  dark wood; and a generation panel in which the phrase smooth white
  foam becomes a vibration waveform labeled sliding vibration
  plus taps. These feed a wide band labeled editable texture layers:
  region plus name plus generated haptic models. Below it, an
  interactive row pairs a screenshot of the browser viewer (relabel,
  repair region, add or delete) with a photograph of the running haptic
  renderer and stylus, joined by paired arrows labeled feel and edit;
  a note says every edit regenerates the layer's haptics live.}
  \label{fig:pipeline}
\end{figure*}

TouchTwin implements a propose--feel--revise workflow
(Figure~\ref{fig:pipeline}). It represents a scene as editable material
layers, each coupling a spatial region and natural-language description
with generated haptic responses. From a phone scan, TouchTwin proposes an
initial layer stack. Users explore it with a haptic stylus and revise
material descriptions or region boundaries; each edit updates the feedback
without exposing low-level signal parameters.

\subsection{Interaction Design and Workflow}
A named material layer is the shared unit manipulated by the automatic
pipeline and the user: a mesh region, material description, and the haptic
models generated from that description. The pipeline initializes these
layers from the scan, but its proposals remain editable. Users therefore
correct the same representation that the system produces rather than
switching to signal-level controls.
Users alternate between feeling and revising the layers. They compare the
rendered feel with a physical scene or an intended result, relabel a surface
that feels wrong, and repair a boundary or add a layer when the spatial
decomposition is wrong. Relabeling invokes the same text-to-haptics path
used for the automatic proposal. Thus, the system might propose
``varnished wood'' and the user replace it with ``rough foam'' without
translating that correction into vibration or friction parameters.
The loop supports replication, in which users correct a draft against
physical objects, and expression, in which they begin with an empty stack.
In both cases, users specify what a surface should feel like and where that
description applies; TouchTwin handles how it is rendered.

\subsection{Creating the Automatic Haptic Draft}
TouchTwin creates its draft from a ${\sim}30$-second phone video. VGGT
recovers camera poses, depth, and a colored point cloud~\cite{wang2025vggt};
a monocular diffusion model completes missing depth~\cite{murre2025}; and
TSDF fusion produces the mesh. We remove unreachable surfaces, repair small
holes, and smooth a copy for haptic interaction. These steps matter because
holes can trap the haptic proxy, disconnected fragments create barriers,
and surface noise can mask synthesized texture. We evaluate the resulting
geometry in \S\ref{sec:eval-recon}.
The system then builds named layers in three steps. SAM~2.1 segments capture
frames; we project their masks onto the mesh and combine observations by
multi-view voting~\cite{ravi2024sam2}. Qwen2.5-VL describes the material and texture
in a clear view of each region~\cite{bai2025qwen25vl}. Finally, the
language-to-haptics model in \S\ref{sec:text2haptic} generates vibration
and tapping responses from each description. The result is deliberately a
draft. Appearance cannot reveal properties such as compliance, so plausible
visual classifications may still feel wrong.

\subsection{Revising the Draft through Language and Spatial Editing}
\begin{figure*}
  \centering
  \includegraphics[width=0.98\textwidth]{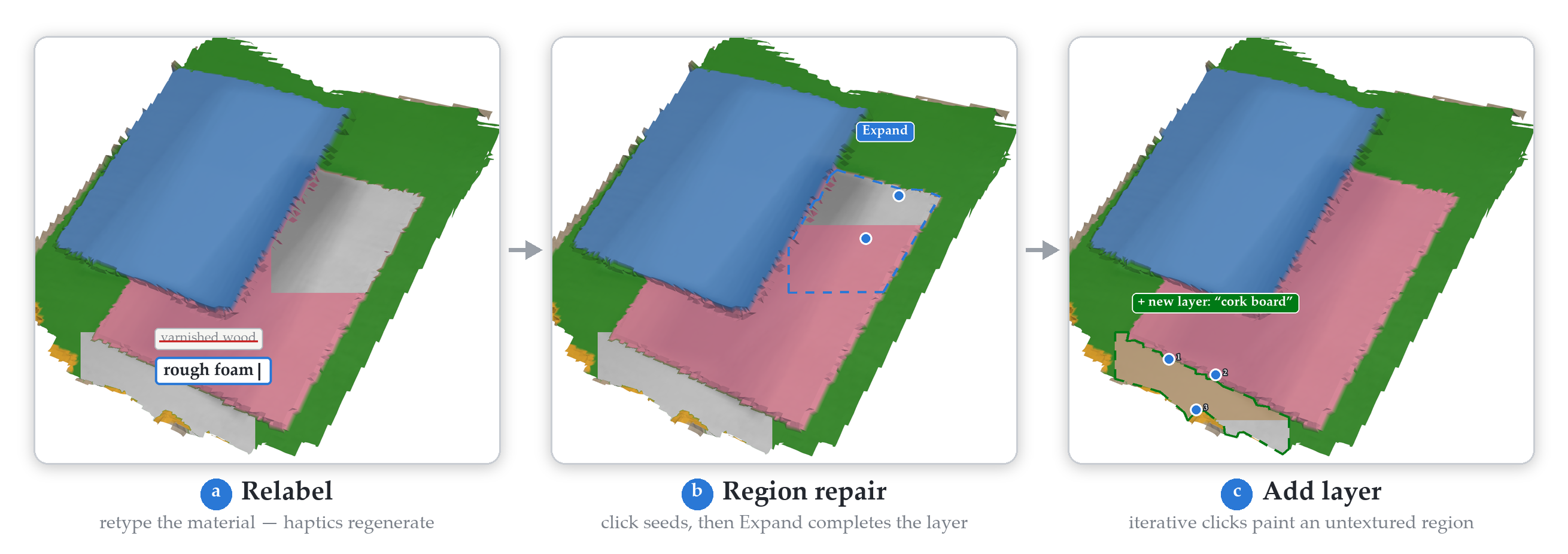}
  \caption{The three correction operations as one progression on a
  study scene. (a)~\emph{Relabel}: the board's proposed name
  ``varnished wood'' is retyped as ``rough foam''; the layer's haptic
  models regenerate as typing settles. At this point the layer is
  still incomplete (grey patch) and a strip of the desk is untextured.
  (b)~\emph{Region repair}: clicks seed the missing patch and
  \emph{Expand} grows them to the layer boundary, completing the foam.
  (c)~\emph{Add layer}: iterative clicks (1--3) progressively paint
  the untextured grey strip as a new ``cork board'' layer.}
  \Description{Three renderings of the same segmented scene mesh, left
  to right. In the first, the pink board layer has a grey unlabeled
  patch and a grey strip of desk below it; the label varnished wood is
  struck through and rough foam is typed as a replacement. In the
  second, two click dots and a dashed boundary cover the grey patch
  next to an Expand button, and half the patch is already pink. In the
  third, the board is fully pink and numbered clicks one to three
  progressively fill the grey desk strip with a tan color under the
  annotation new layer: cork board.}
  \label{fig:operations}
\end{figure*}

The Viser-based browser interface~\cite{yi2025viser} provides three
operations corresponding to semantic, boundary, and omission errors:
relabel, region repair, and add/delete (Figure~\ref{fig:operations}).
\textbf{Relabel} corrects a region whose material feels wrong. The user
selects its layer and types a new name or description, such as replacing
``varnished wood'' with ``rough foam'' (Figure~\ref{fig:operations}a).
TouchTwin regenerates the haptic models after typing settles and updates the
running renderer without restarting the scene.
\textbf{Region repair} corrects a layer that covers too much or too little
surface. Users seed faces, then a SAM-assisted selector expands or shrinks
the selection using visual similarity, the material description, and mesh
adjacency (Figure~\ref{fig:operations}b). They can also transfer individual
faces between neighboring layers. \textbf{Add/delete} creates a named layer
for an untextured region or removes one and returns its faces to the base
(Figure~\ref{fig:operations}c).
Color overlays show each layer's extent, and the haptic display identifies
the material under the stylus. The same operations support freeform
authoring from an empty stack.

\subsection{Language as the Haptic Authoring Interface}
\label{sec:text2haptic}
Natural-language descriptions connect author intent to the rendered signals.
Users can enter material names such as ``cardboard'' or richer descriptions
such as \emph{gritty but cushioned}. VLM proposals and user descriptions
follow the same text-to-haptics path, so a correction directly replaces the
AI's semantic proposal.
We use Qian et al.'s language-guided texture model~\cite{qian2026language}.
It embeds a description with CLIP ViT-L/14~\cite{radford2021learning} and
maps the embedding into a latent space learned from tool--surface
interactions with 100 measured materials~\cite{culbertson2014hundred}. The
decoder generates force- and speed-dependent sliding vibrations and
speed-dependent tap responses. Thus, open descriptions produce combinations
of dynamics learned from measured contact rather than arbitrary signals.

\subsection{Haptic Rendering and Implementation}
\label{sec:rendering}
\begin{figure}
  \centering
  \includegraphics[width=0.68\textwidth]{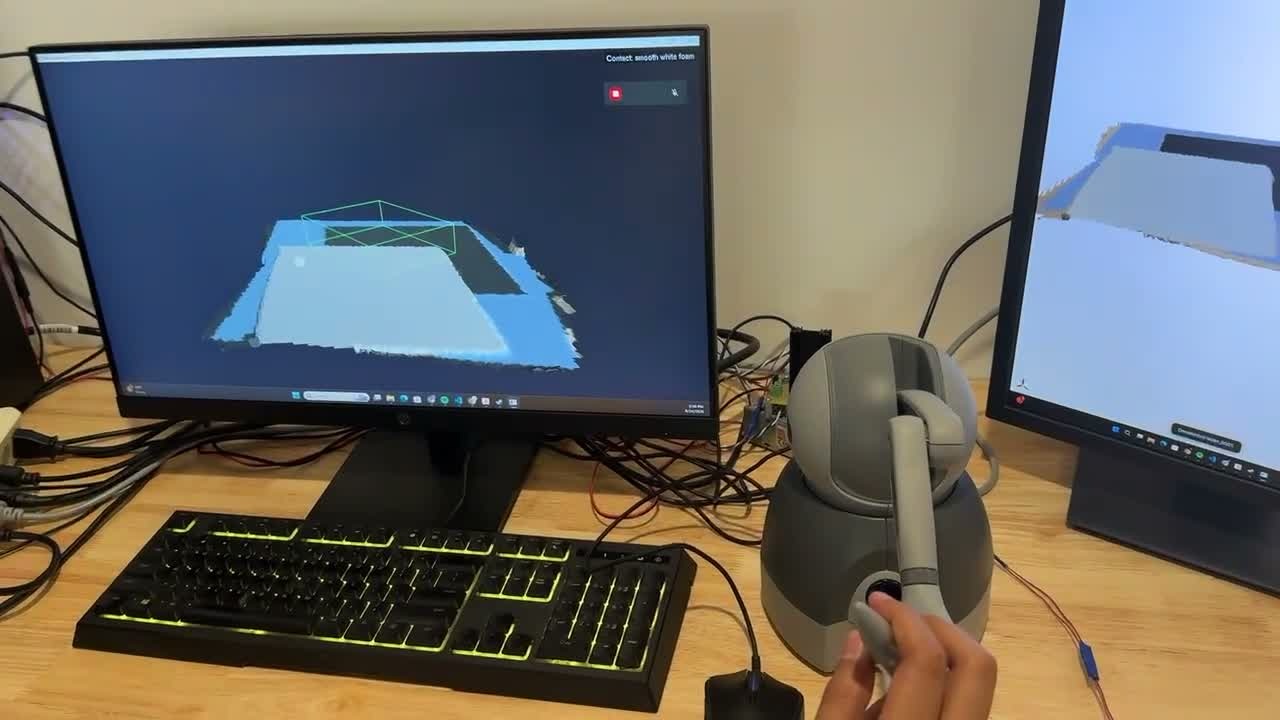}\\[3pt]
  \includegraphics[width=0.255\textwidth]{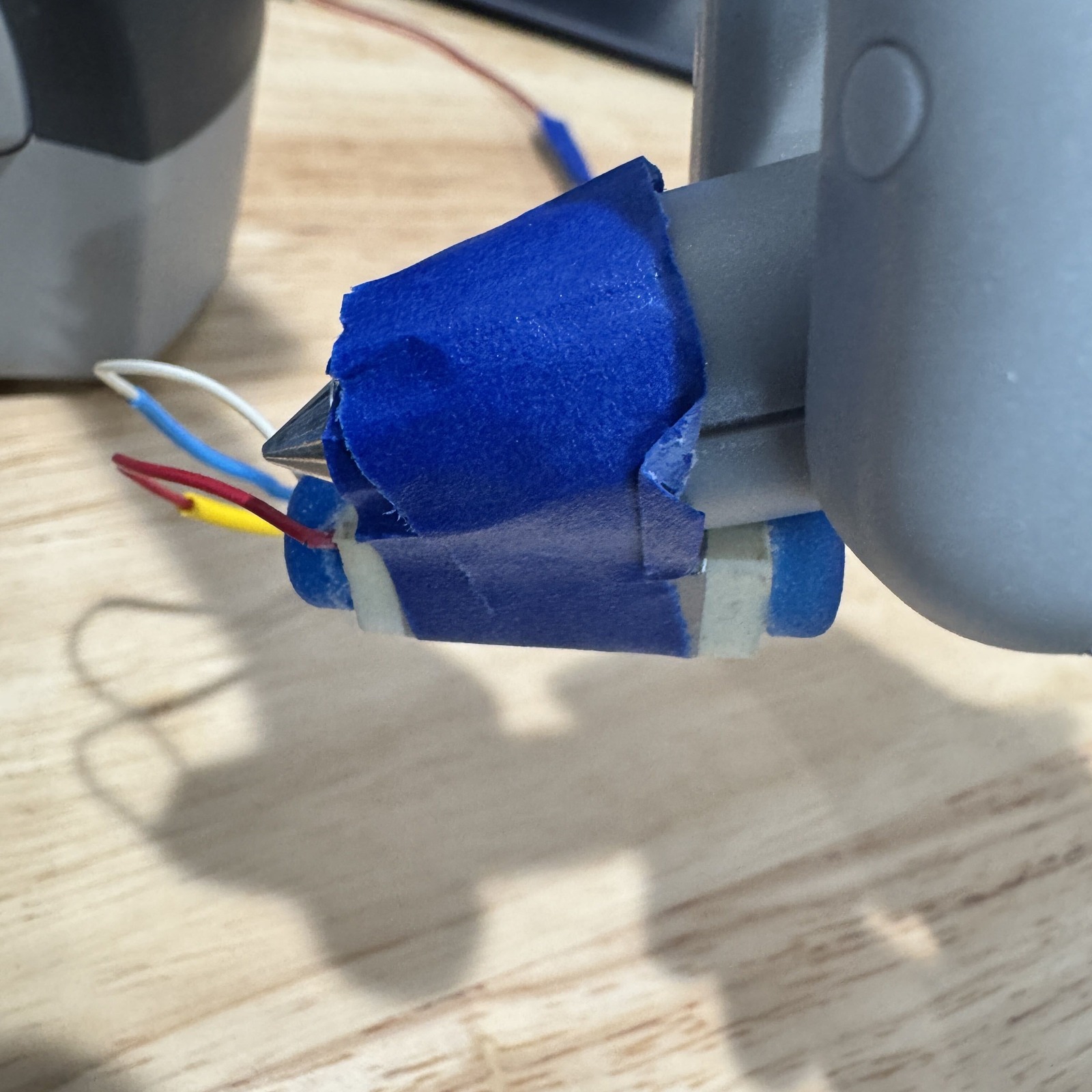}\hspace{0.012\textwidth}%
  \includegraphics[width=0.413\textwidth]{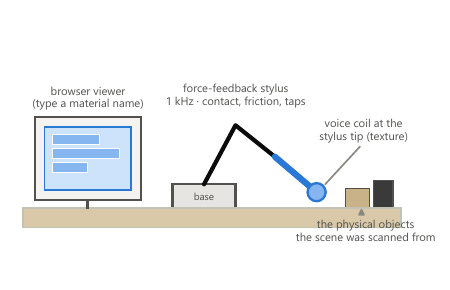}
  \caption{Study apparatus. Top: the setup in use; the haptic renderer
  with the contact-material HUD on the main display (here reading
  ``smooth white foam''), the authoring viewer on the adjacent monitor,
  and the stylus under the operator's hand. Bottom left: the MM3C
  Haptuator voice-coil actuator (TactileLabs) mounted at the tip of the
  force-feedback stylus, so the texture signal is delivered at the
  simulated contact point. Bottom right: schematic of the study
  arrangement, in which the physical scanned objects sit on the same
  desk, within reach.}
  \Description{Three-panel figure of the study setup. The top panel is
  a photograph of the apparatus in use: a monitor showing the haptic
  renderer with a contact material name in the corner, a second monitor
  with the authoring viewer, a keyboard, and a hand holding the stylus
  of the grounded force-feedback device. The bottom-left panel is a
  close-up photograph of the MM3C Haptuator voice-coil actuator mounted
  at the stylus tip with its drive wires visible. The bottom-right panel
  is a line-drawing schematic of the study arrangement in side view,
  with the browser viewer, the articulated stylus with the coil at its
  tip, and the scanned physical objects labeled.}
  \label{fig:apparatus}
\end{figure}

TouchTwin uses a grounded Phantom Touch with 3-DoF force feedback (3D
Systems) and an MM3C Haptuator voice-coil actuator (TactileLabs) mounted
near the stylus tip (Figure~\ref{fig:apparatus}). Following the HaTT
framework~\cite{culbertson2014hundred,culbertson2014modeling}, the Phantom
renders contact, friction, and taps at a 1\,kHz servo rate; the Haptuator
renders sliding-texture vibration.

At each contact, the renderer identifies the active layer on the smoothed
haptic mesh. A proxy algorithm renders stiff contact and the selected
material's measured friction coefficient. Impacts trigger speed-dependent
tap transients. During sliding, an autoregressive texture model conditioned
on tangential hand speed and contact force produces vibration above a
5\,mm/s threshold. The Haptuator places this signal near the simulated
contact point, avoiding attenuation through the grounded device's linkage.
The 1\,kHz signal is sent over a 48\,kHz audio channel with ${\sim}1$\,ms
buffers and ${\sim}10$\,ms reported output latency. Appendix~\ref{app:params}
reports renderer parameters.

Updated haptic models are generated outside the real-time servo loop and
published atomically, preventing generation from blocking the 1\,kHz
contact loop or exposing a partial update. Users can therefore feel a
revised material without restarting the scene.

\section{Characterizing the Automatic Haptic Draft}
\label{sec:techeval}

Before studying how people revise TouchTwin's drafts, we characterize the
starting point itself: whether its geometry supports haptic exploration,
what material errors it contains, and how much correction remains.
The workflow does not require a correct final scene, but it does require a
draft that users can explore and meaningfully revise.

We built six scenes from the 100-material library, using the same product
when available and one from the same category otherwise. Each scene contained
four ground-truth (GT) surfaces (24 total). For scoring, we resolve each
open-ended VLM description to its nearest library material in the model's
language-aligned latent space. We report exact-material accuracy, accuracy
within ten author-defined families (wood, metal, plastic, foam/rubber,
leather, fabric, paper, ceramic/stone, abrasive, and carbon). We also use two
signal-space distances. \emph{Assignment distance} compares the measured
models of the resolved and GT materials, isolating the material assignment.
\emph{Generated-model distance} compares the response generated from a
description with the measured GT model, incorporating both assignment and
generation error. Both are cosine distances between L2-normalized vectors of
AR line-spectral frequencies and tap envelopes. Lower is better: identical
models score 0, while random library pairs average ${\approx}0.84$.

\subsection{Haptic Usability of the Reconstructed Geometry}
\label{sec:eval-recon}
A visually acceptable mesh may still impede touch: holes can trap the haptic
proxy, disconnected fragments create barriers, and surface noise can mask
texture feedback. We therefore assessed geometric properties that directly
affect stylus interaction.
Compared with Poisson reconstructions of the same VGGT point clouds,
TouchTwin reduced boundary-defect edges from $932$--$2{,}483$ to 0--4 and
components from as many as 229 to 1--2 (Table~\ref{tab:recon}). All intended
surfaces were proxy-reachable; residual planar warp was $0.4$--$1.8\%$, with
near-zero median dihedral roughness after smoothing. This does not establish
a generally superior reconstruction method, but verifies a usable substrate
for these near-planar scenes.

\begin{table}[t]
  \caption{Reconstruction quality per scene: boundary-defect edges /
  connected components, for our pipeline vs.\ Poisson reconstruction of
  the same point cloud (lower is better for both).}
  \label{tab:recon}
  \begin{tabular}{lrr}
    \toprule
    Scene & Ours (defects/comp.) & Poisson (defects/comp.)\\
    \midrule
    S1 & 3 / 1 & 1{,}471 / 22\\
    S2 & \textbf{0} / 1 & 2{,}483 / 229\\
    S3 & 3 / 2 & 1{,}236 / 19\\
    S4 & 4 / 1 & 932 / 12\\
    S5 & 4 / 1 & 1{,}298 / 23\\
    S6 & \textbf{0} / 1 & 1{,}045 / 5\\
    \bottomrule
  \end{tabular}
\end{table}

\subsection{Quality and Failure Modes of the Automatic Proposal}
\label{sec:eval-what}
Segmentation produced 4--7 regions per scene, which resolved to four material
layers. The VLM identified the exact GT material for 13 of 24 surfaces
(54\%; 25--100\% across scenes), with 58\% material-family accuracy.
Table~\ref{tab:what} also reports face-weighted accuracy and material-region
IoU. With only four GT surfaces per scene, IoU largely tracks material
accuracy; boundary evaluation would require per-face ground truth.
Several visually plausible errors mapped compliant materials to rigid ones,
including carpet to wood, foam to carbon fiber, and velvet to wood. Such
properties are unavailable from appearance but perceptible to a user with
the physical referent. Assignment distance reached 0.50 on the worst scene
(random pairs: ${\approx}0.84$), indicating consequential room for revision.

\begin{table}[t]
  \caption{Automatic proposal vs.\ material ground truth (known by
  construction), per scene: top-1 / face-weighted / family accuracy (\%),
  material-region IoU, assignment distance between the resolved and
  true material (random-pair baseline ${\approx}0.84$), and the number
  of relabel corrections needed to reach a perfect scene.}
  \label{tab:what}
  \begin{tabular}{lrrrrrr}
    \toprule
    Scene & Top-1 & Face-wt & Family & IoU & Assign. & Corr.\\
    \midrule
    S1 & 75 & 76 & 75 & 0.67 & 0.07 & 1\\
    S2 & 25 & 24 & 25 & 0.25 & 0.43 & 3\\
    S3 & 25 & 15 & 50 & 0.25 & 0.50 & 3\\
    S4 & 50 & 35 & 50 & 0.50 & 0.35 & 2\\
    S5 & \textbf{100} & \textbf{100} & \textbf{100} & \textbf{1.00} & \textbf{0.00} & \textbf{0}\\
    S6 & 50 & 28 & 50 & 0.50 & 0.33 & 2\\
    \bottomrule
  \end{tabular}
\end{table}

\begin{figure}
  \centering
  \includegraphics[width=0.62\columnwidth]{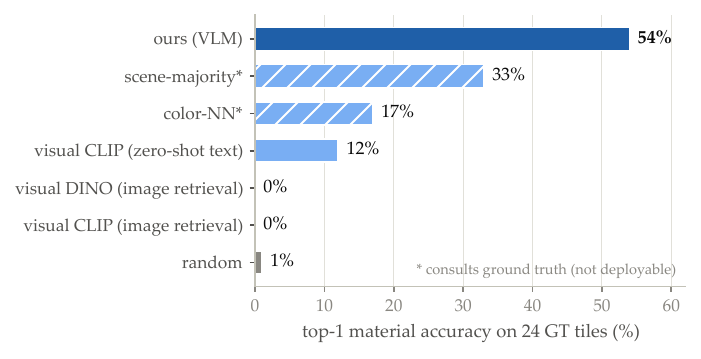}
  \caption{Average top-1 material accuracy on the 24 GT surfaces. Hatched
  baselines consult ground truth and are not deployable. Image-to-exemplar
  retrieval's 0\% is not on its own evidence of failure (24 surfaces over
  100 classes makes 0 hits the likeliest chance outcome); the fairer
  visual baseline, zero-shot text classification against the 100
  material names with the same frozen CLIP encoder, clears chance at
  12\% but stays 42 points behind the VLM pipeline.}
  \Description{Horizontal bar chart of top-1 accuracy on 24 ground-truth
  surfaces, best to worst: ours using the VLM 54 percent; scene-majority
  33 percent and color nearest-neighbor 17 percent, both hatched as
  ground-truth-aware; visual CLIP zero-shot text classification 12
  percent; visual CLIP and visual DINO image-to-exemplar retrieval 0
  percent each; and random at 1 percent.}
  \label{fig:baselines}
\end{figure}

Zero-shot CLIP classification against the same 100 names reached 12\%
exact-material and 29\% family accuracy, versus 54\% and 58\% for the VLM
(Figure~\ref{fig:baselines}). Frozen CLIP and DINO image-to-exemplar
retrieval produced no exact matches, which is consistent with chance for
24 surfaces and 100 classes. GT-dependent scene-majority and color-matching
referents reached 33\% and 17\% exact accuracy. We treat this small
benchmark as descriptive: it characterizes useful but imperfect assistance,
not a new material-recognition method.

\subsection{Correction Required from the Automatic Draft}
\label{sec:eval-effort}
Reaching the GT assignments would require 0--3 relabels per four-surface
scene (Table~\ref{tab:what}). These arithmetic counts assume known GT and
are not measured effort; the study reports observed behavior. On four
scenes, pairwise assignment distances among proposed materials ranged from
$0.24$--$0.57$. The other two legitimately contained two wood regions, so
not every pair should differ.
The automatic output is therefore a structured but imperfect draft. Its
geometry supports exploration, while its material errors leave a small
number of potentially consequential corrections. The user study tests
whether people can make those corrections and improve the experience.

\section{User Study: Evaluating Propose--Feel--Revise Haptic Authoring}
\label{sec:study}


We conducted a within-subjects study of whether non-experts find an
automatic haptic draft useful, can improve it through semantic and spatial
edits, and can use the same mechanism for freeform authoring.

\subsection{Study Goals and Design}
\begin{figure*}
  \centering
  \setlength{\tabcolsep}{1.5pt}
  \newcommand{\scenecell}[1]{\includegraphics[width=0.152\textwidth,height=0.125\textwidth,keepaspectratio]{#1}}
  \begin{tabular}{c@{\hspace{4pt}}cccccc}
    & {\scriptsize\shortstack{S1 · before/after\\(easier) · 75\%}}
    & {\scriptsize\shortstack{S2 · reserve\\· 25\%}}
    & {\scriptsize\shortstack{S3 · reserve\\· 25\%}}
    & {\scriptsize\shortstack{S4 · freeform\\canvas · 50\%}}
    & {\scriptsize\shortstack{S5 · training\\· 100\%}}
    & {\scriptsize\shortstack{S6 · before/after\\(harder) · 50\%}}\\
    \rotatebox{90}{\scriptsize photo} &
    \scenecell{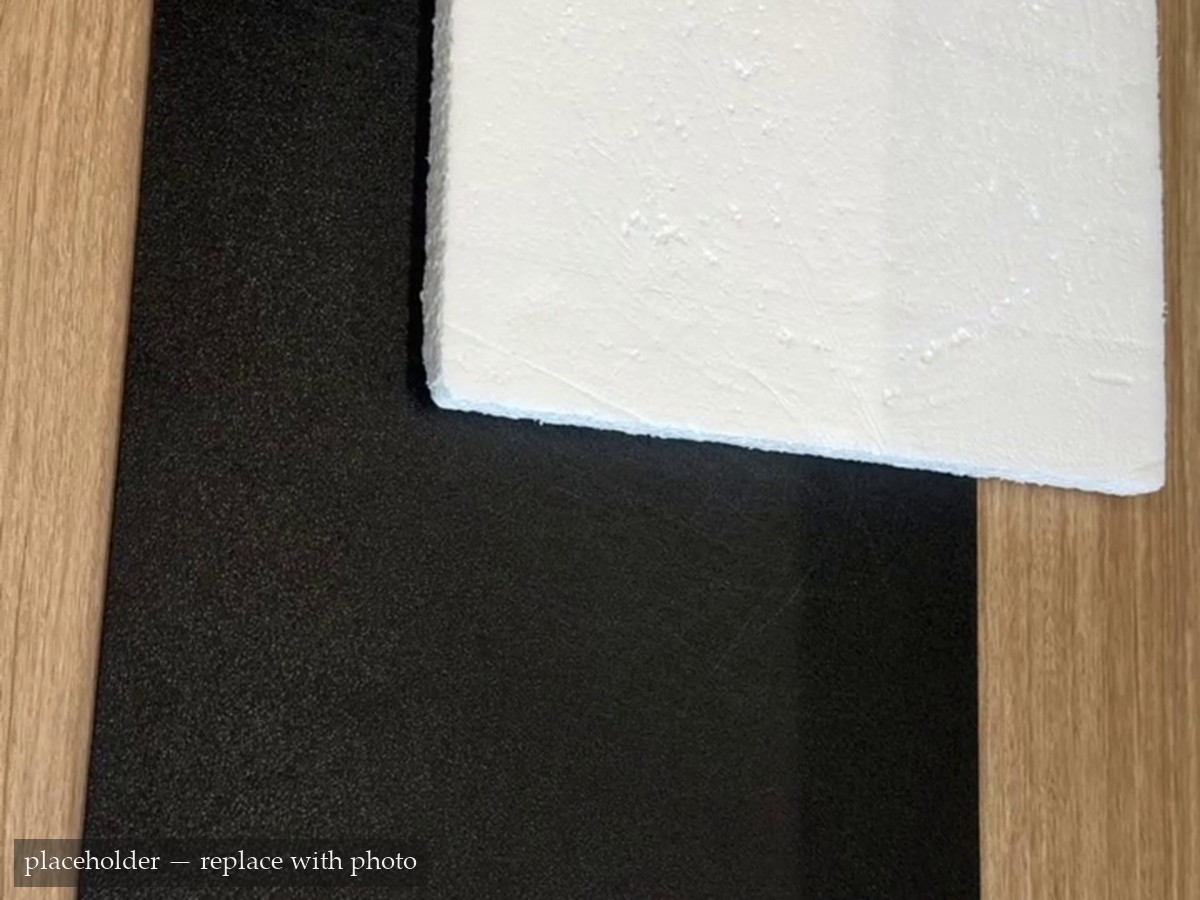} & \scenecell{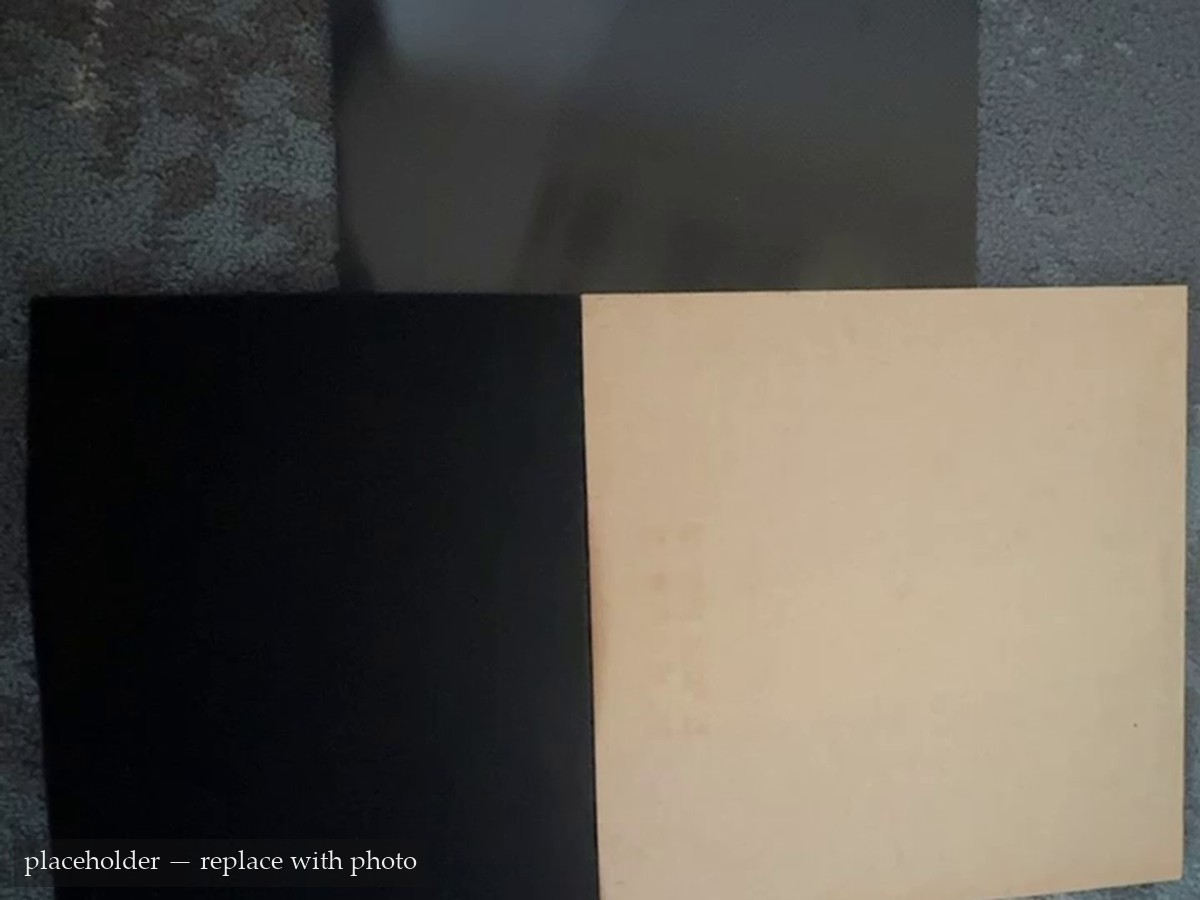} &
    \scenecell{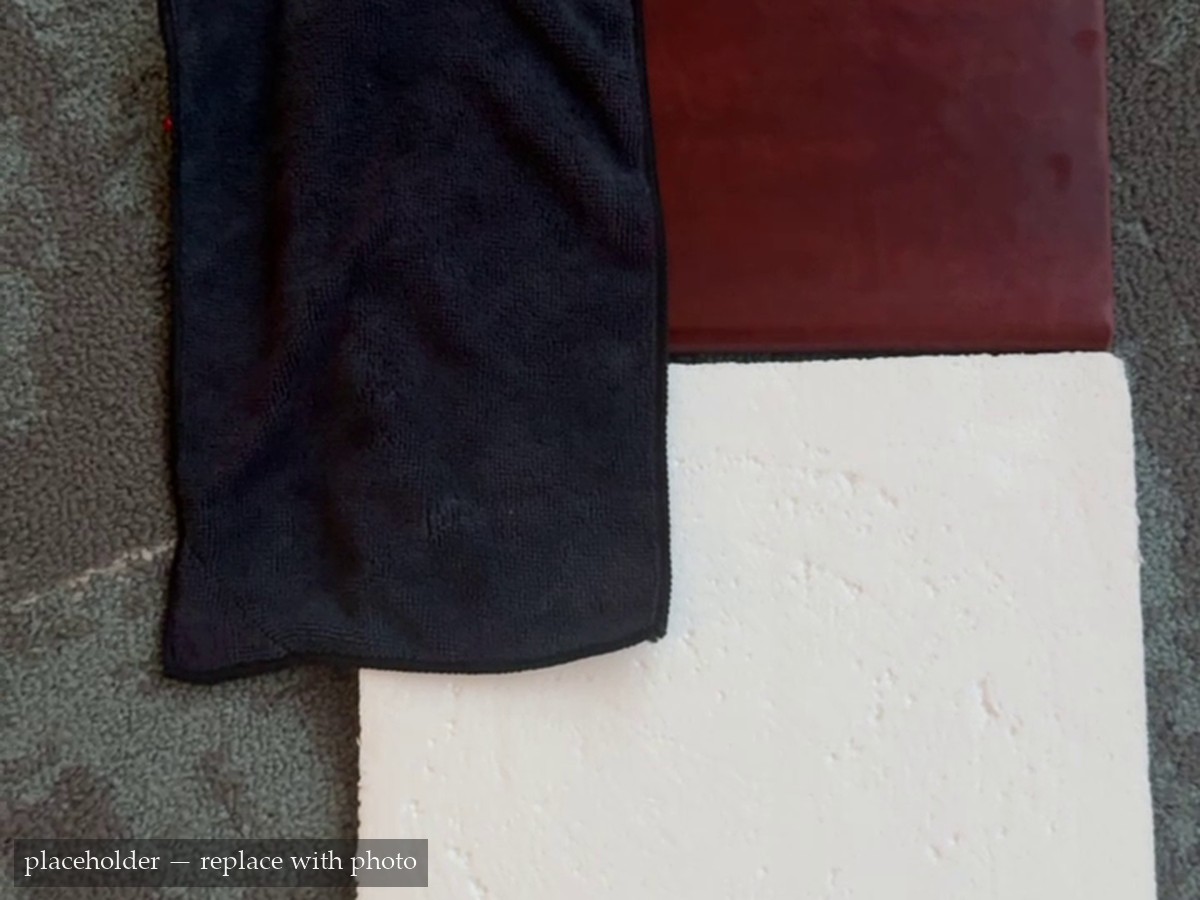} & \scenecell{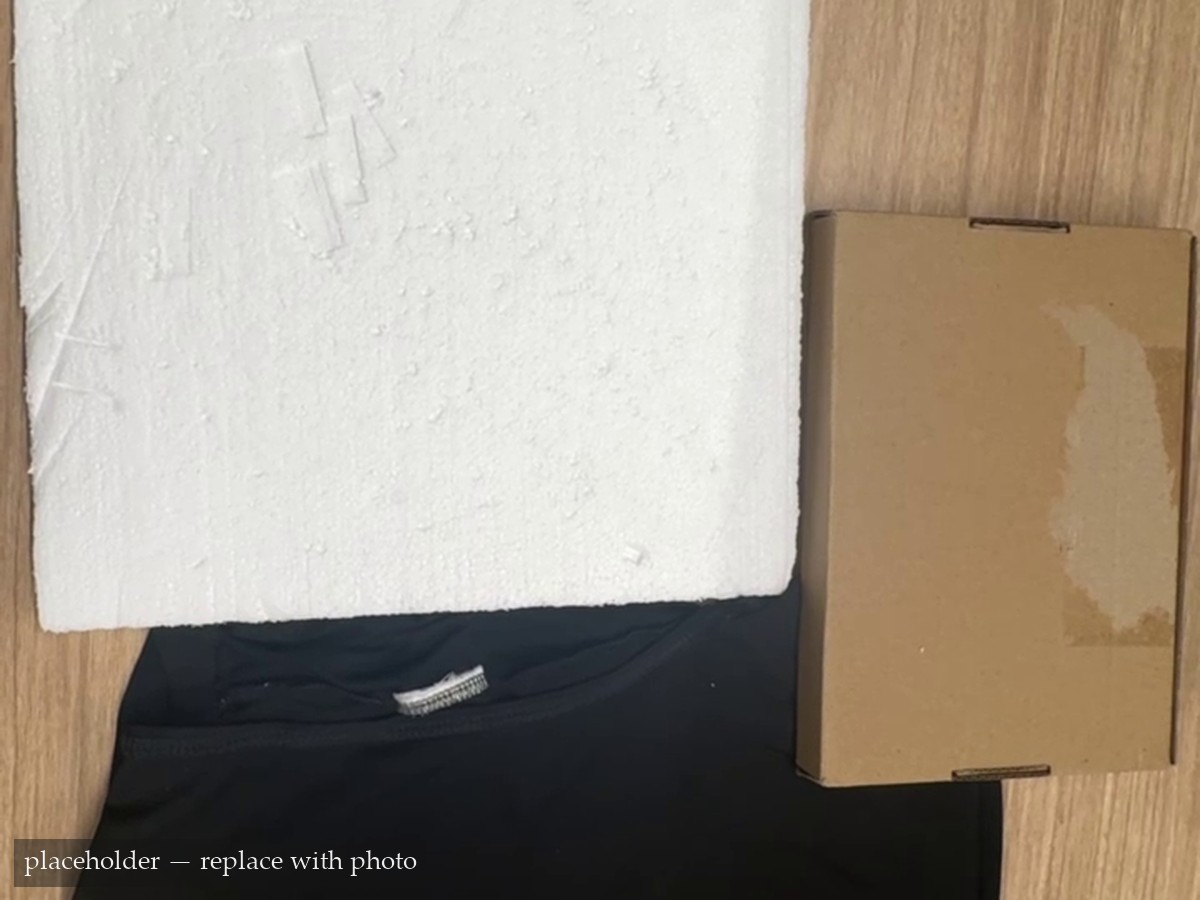} &
    \scenecell{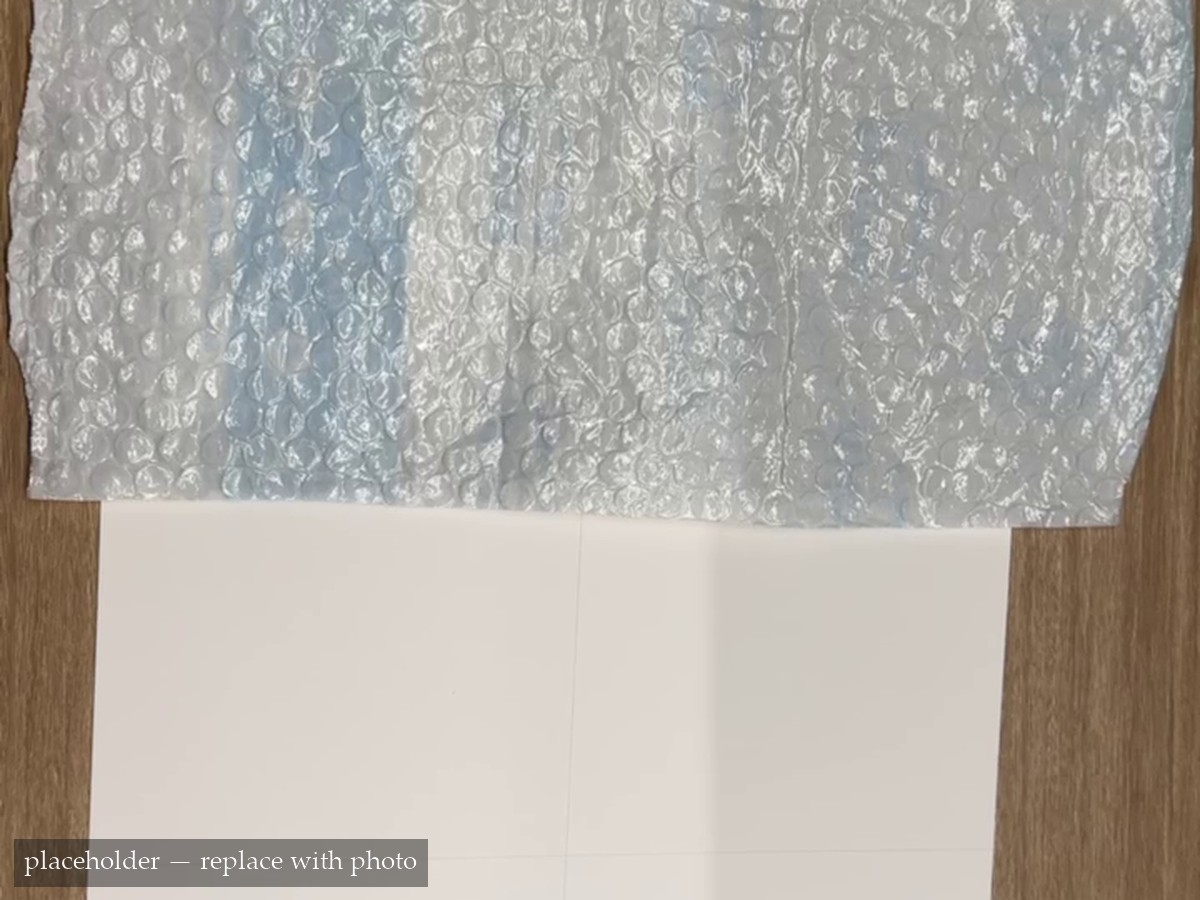} & \scenecell{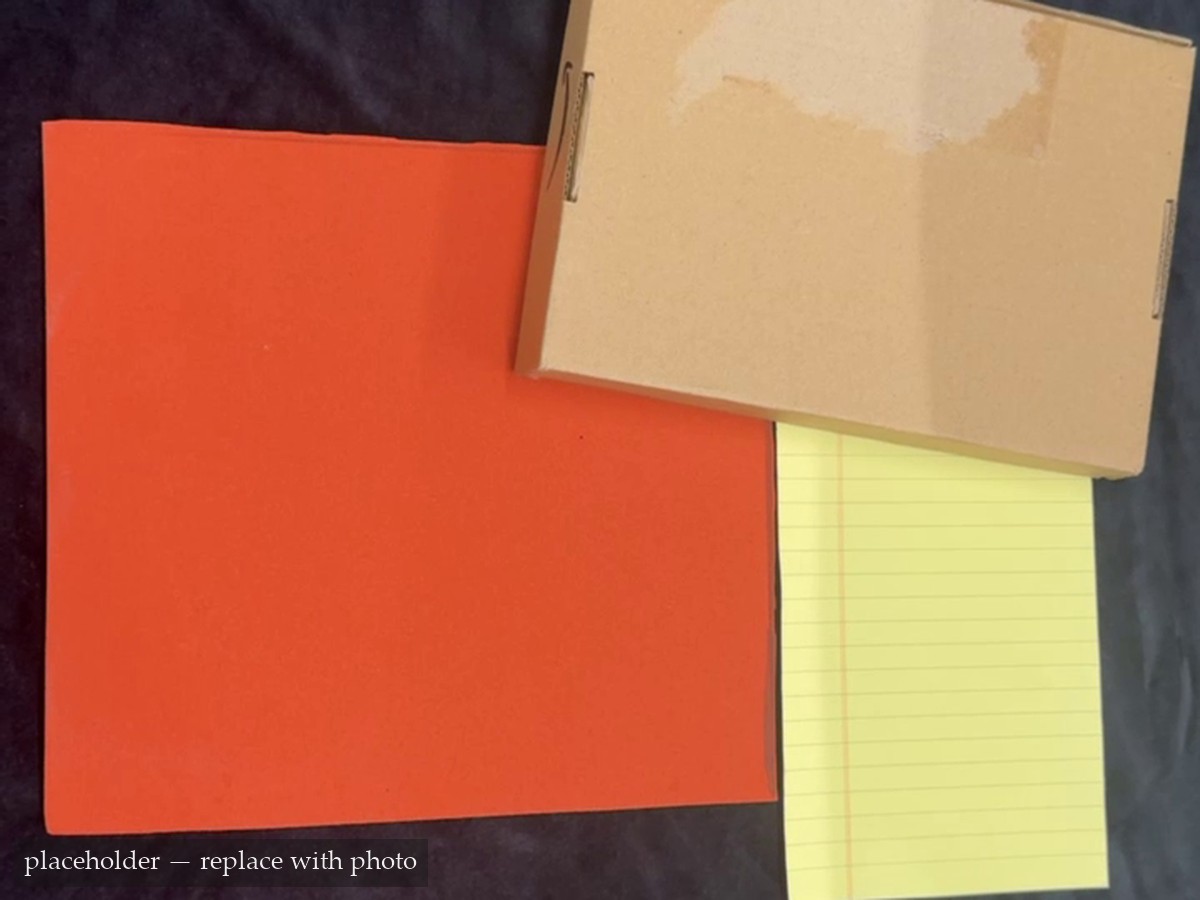}\\
    \rotatebox{90}{\scriptsize reconstruction} &
    \scenecell{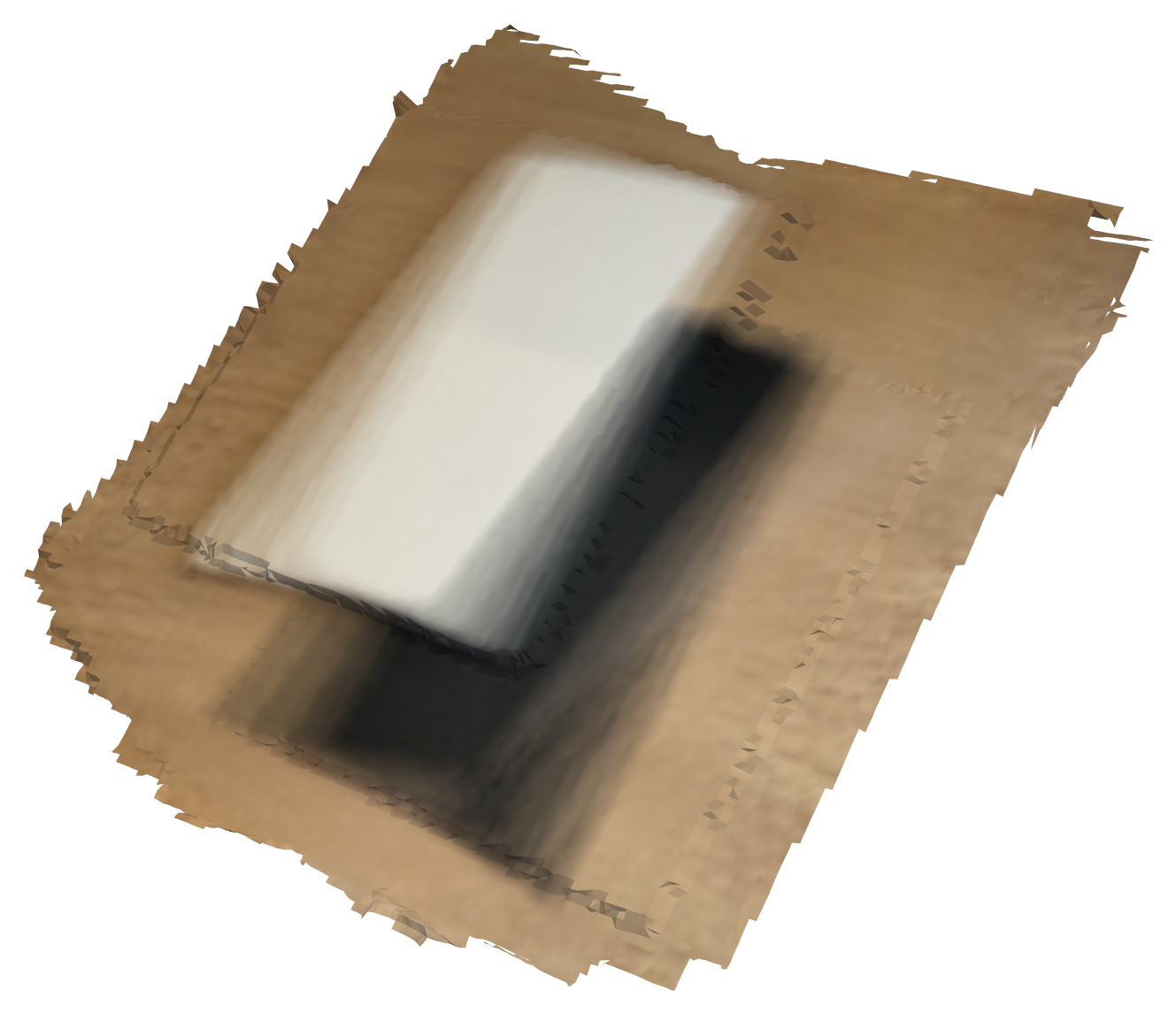} & \scenecell{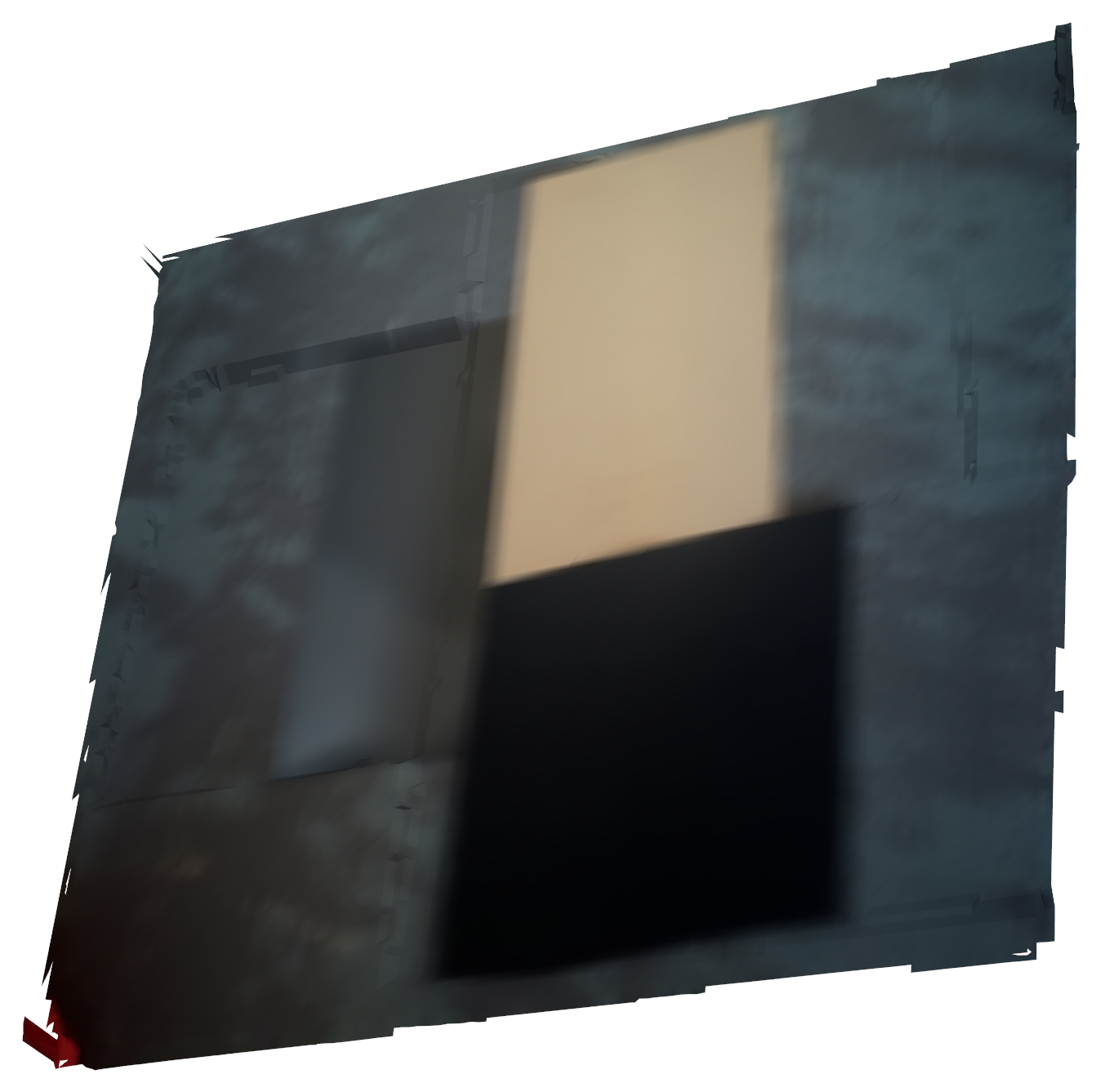} &
    \scenecell{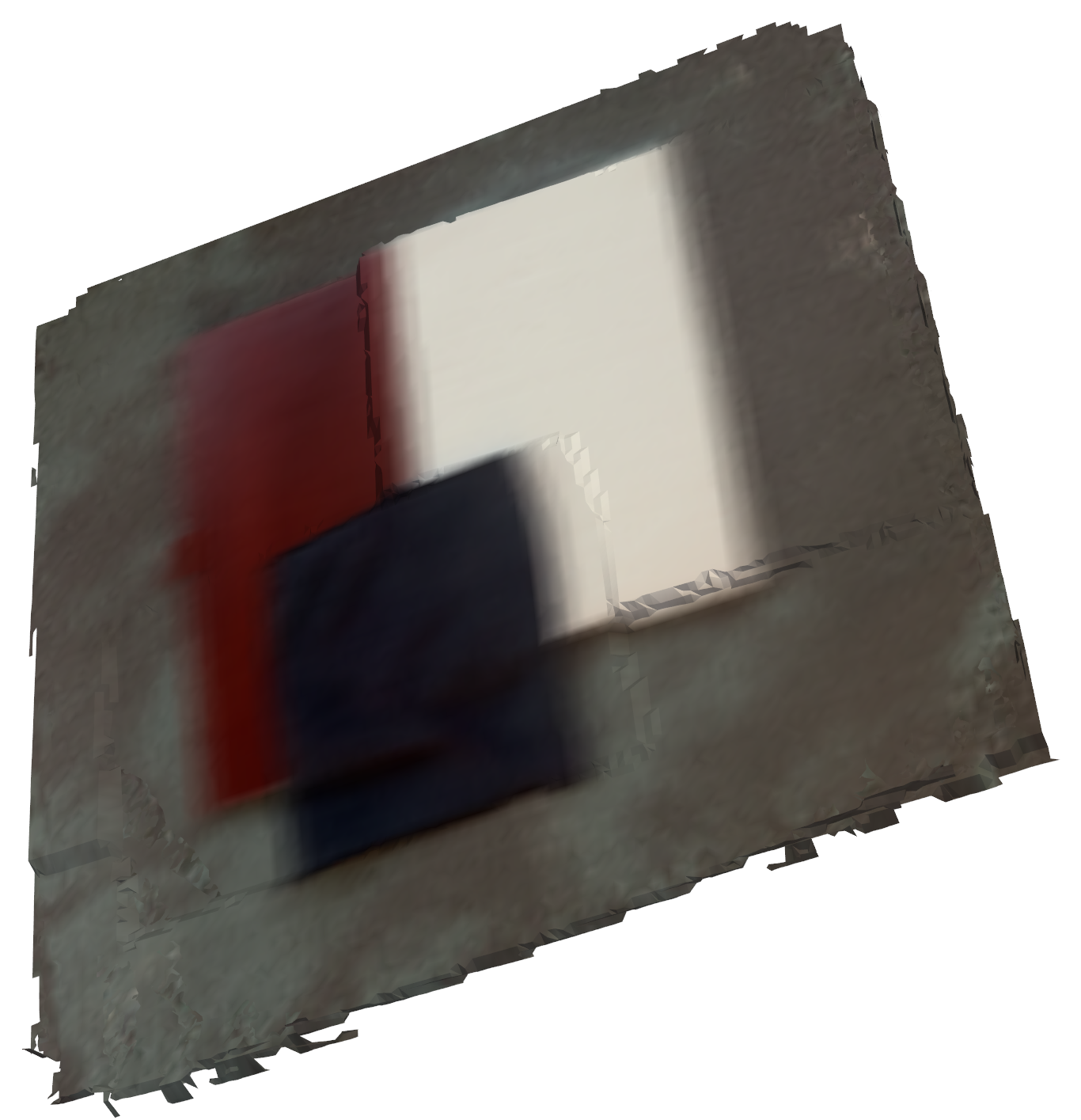} & \scenecell{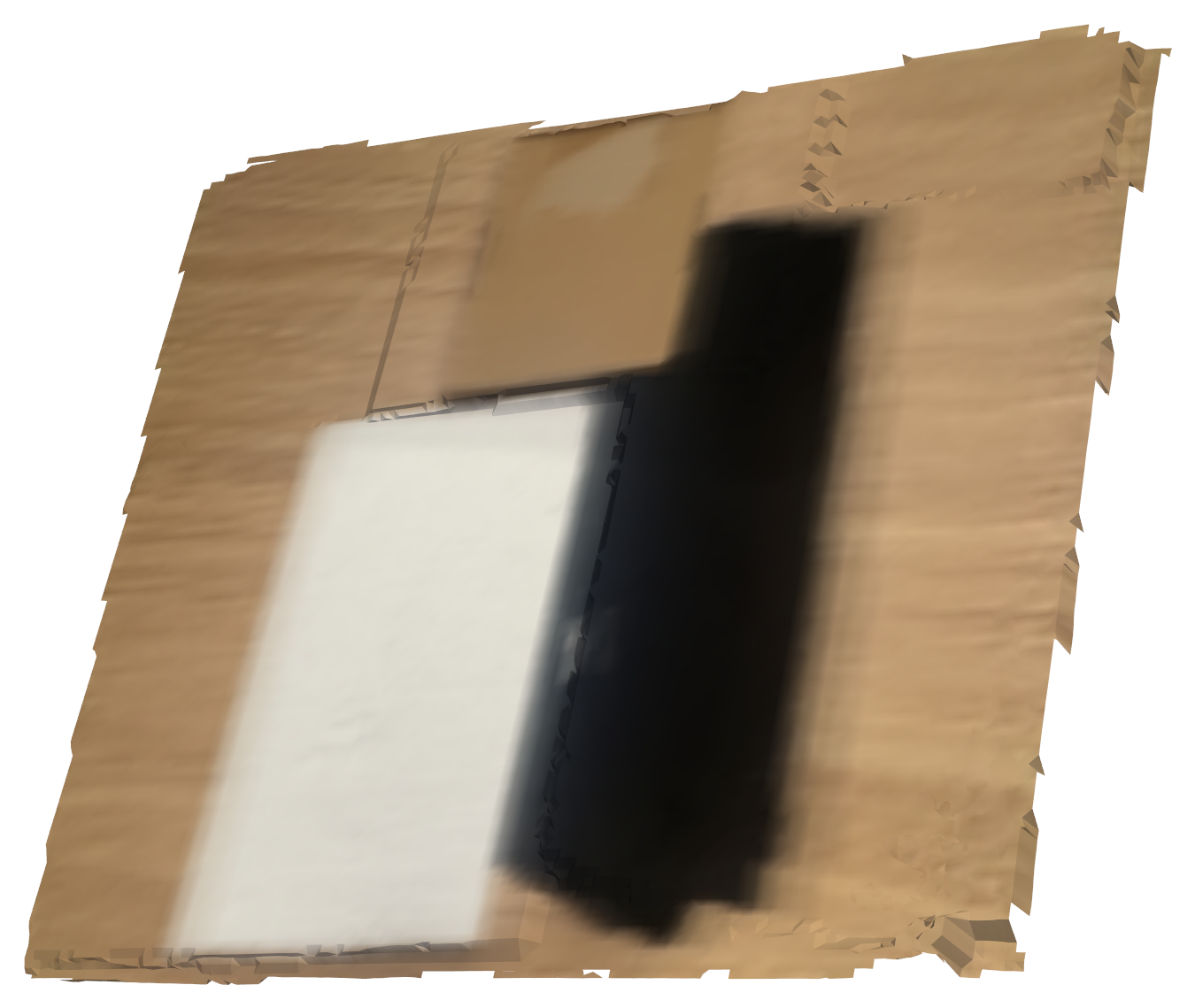} &
    \scenecell{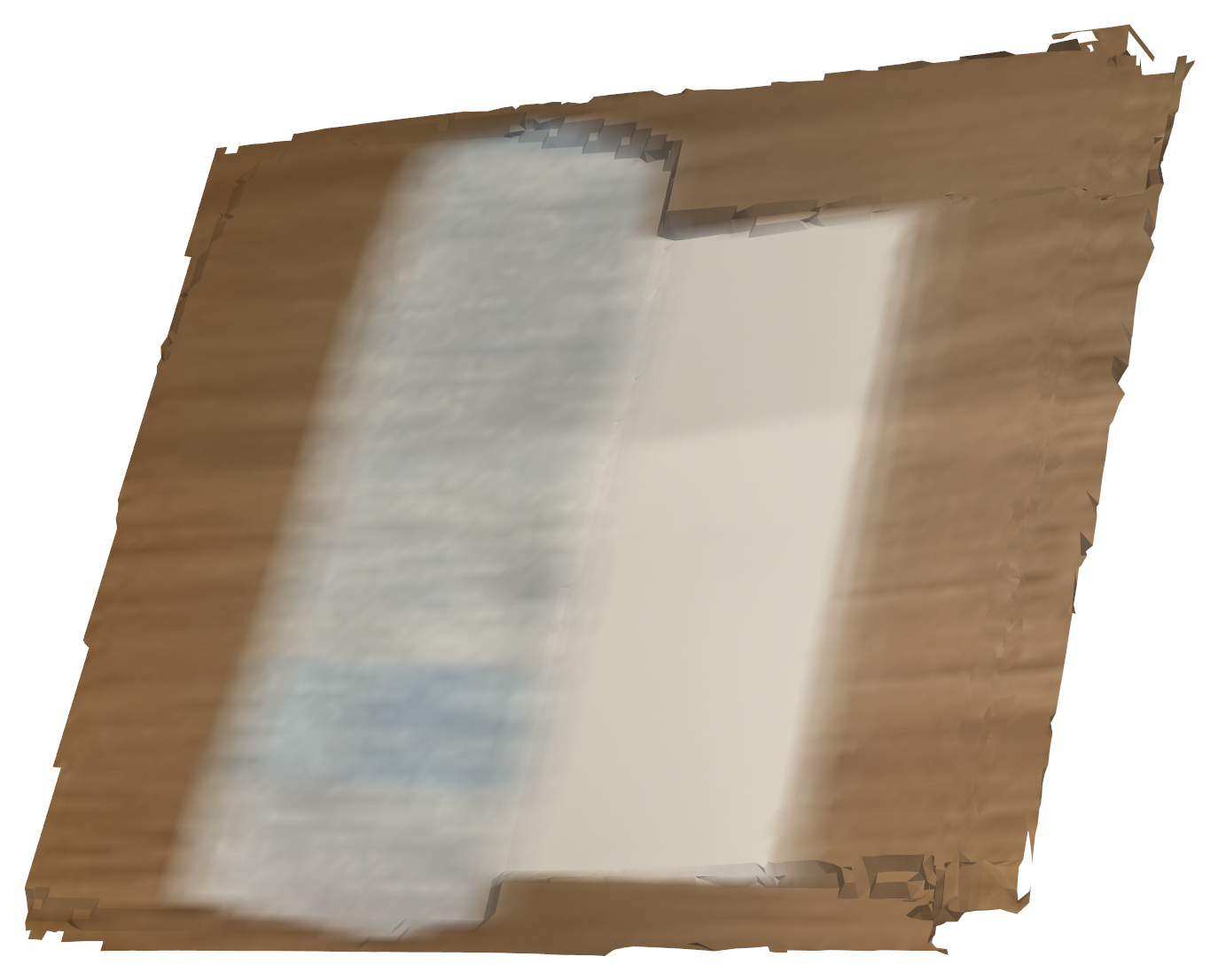} & \scenecell{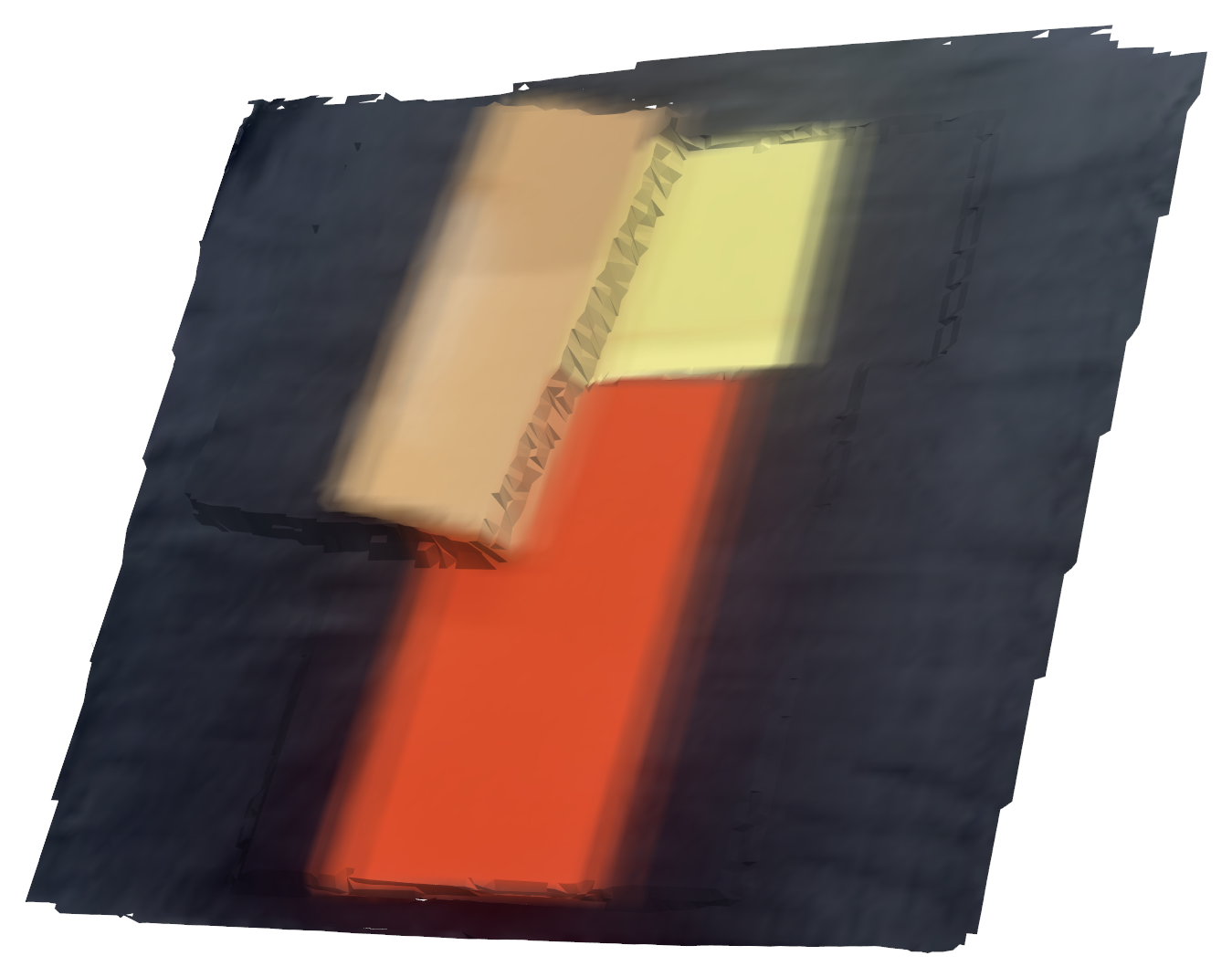}\\
    \rotatebox{90}{\scriptsize auto proposal} &
    \scenecell{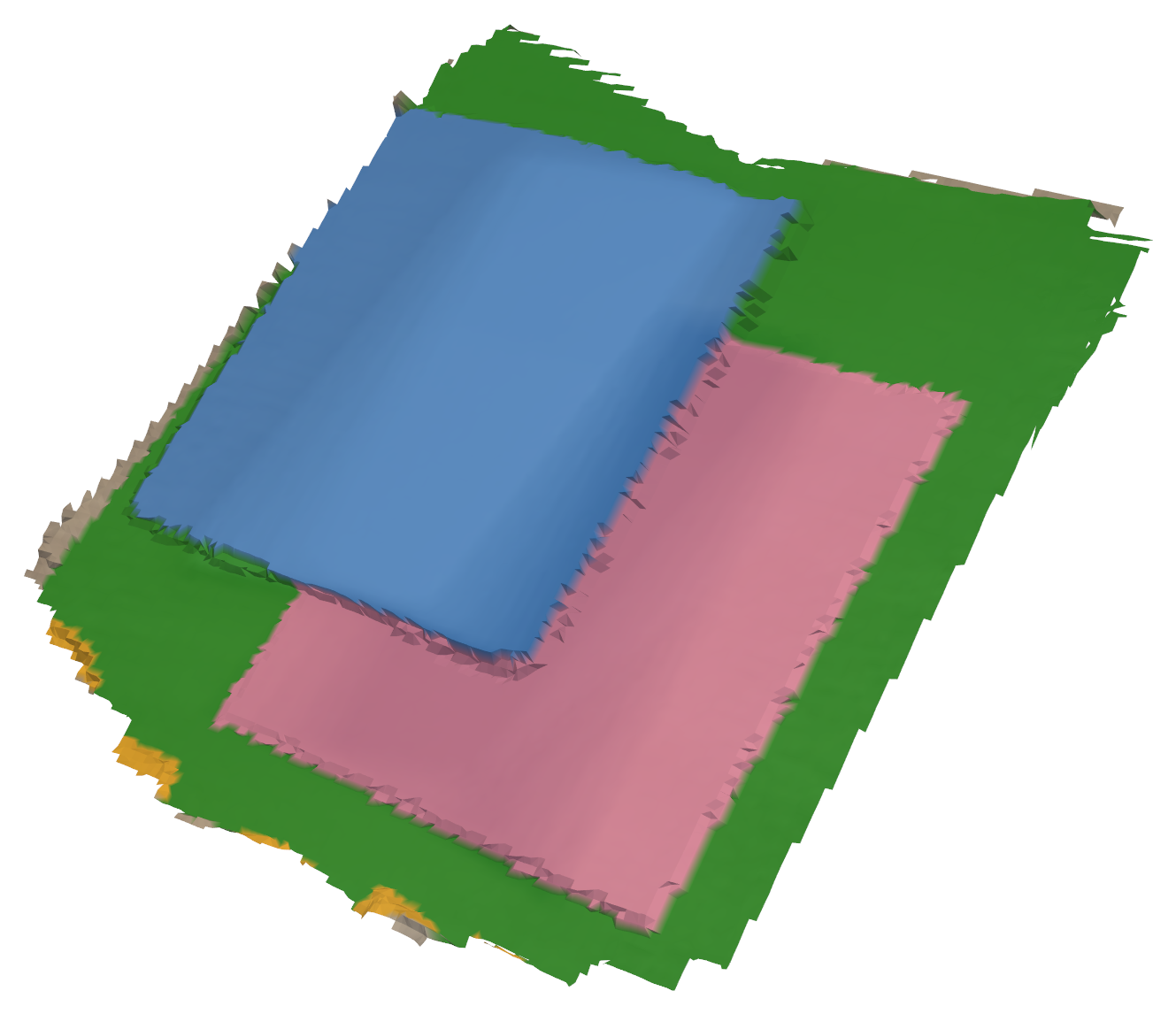} & \scenecell{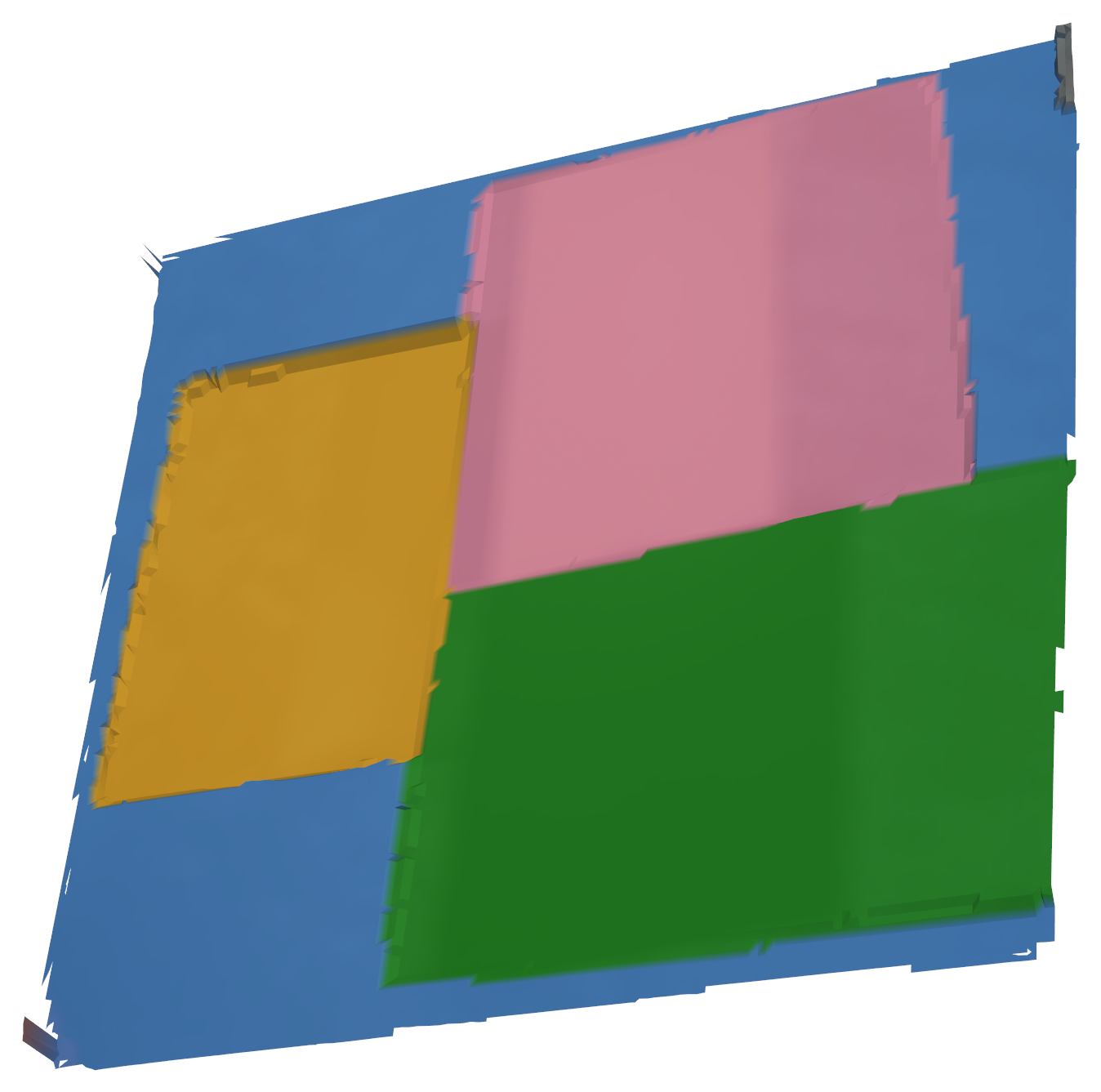} &
    \scenecell{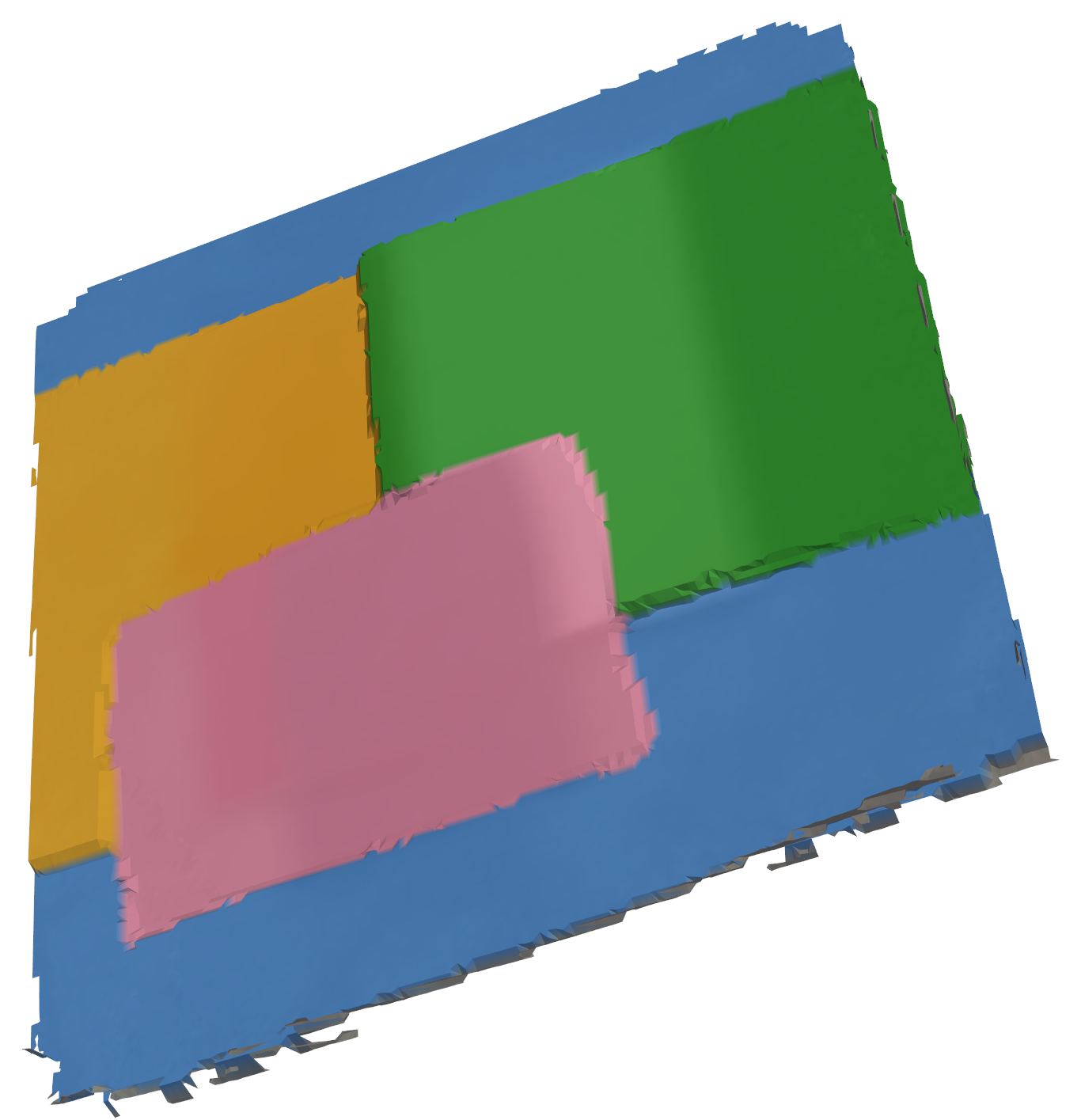} & \scenecell{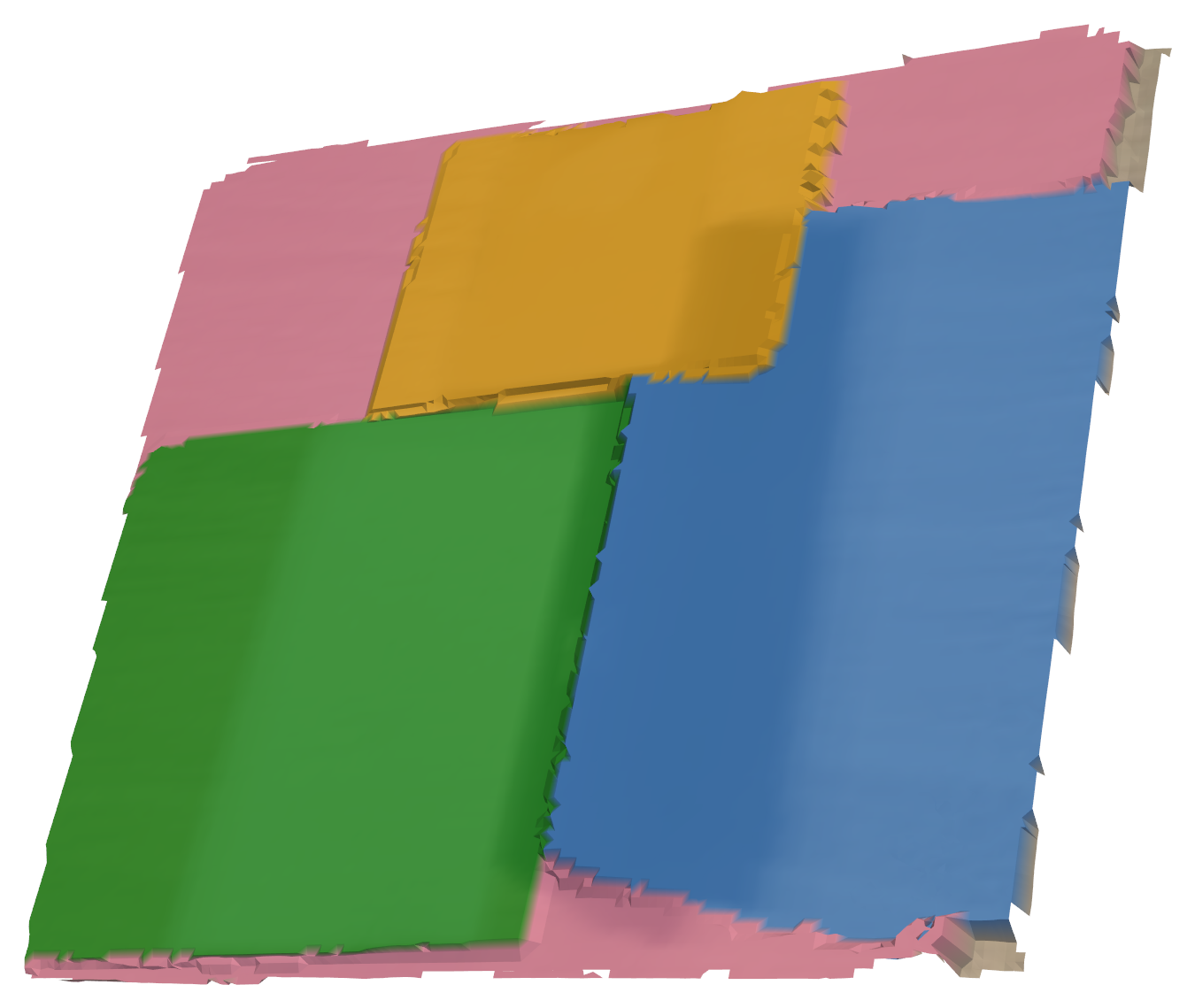} &
    \scenecell{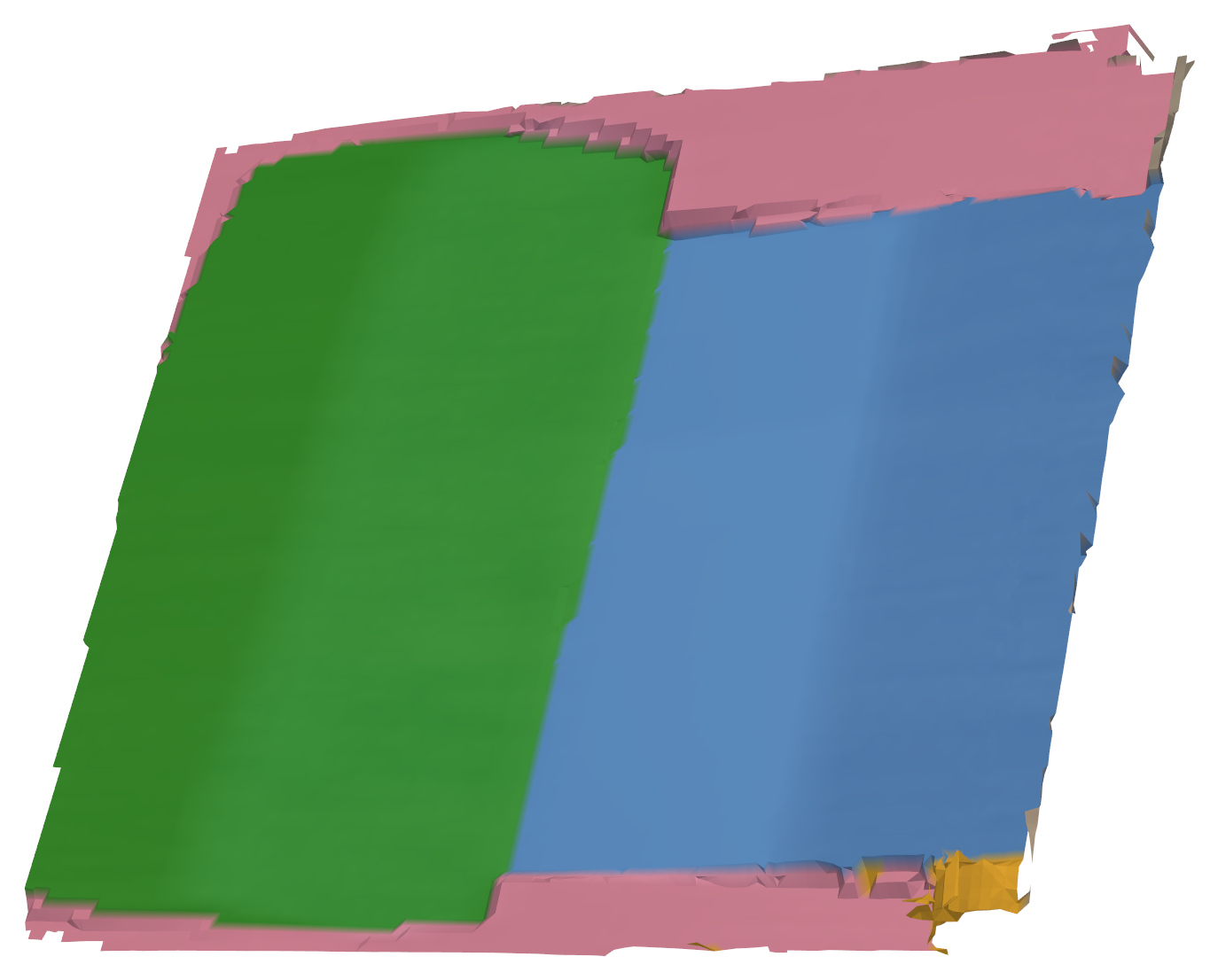} & \scenecell{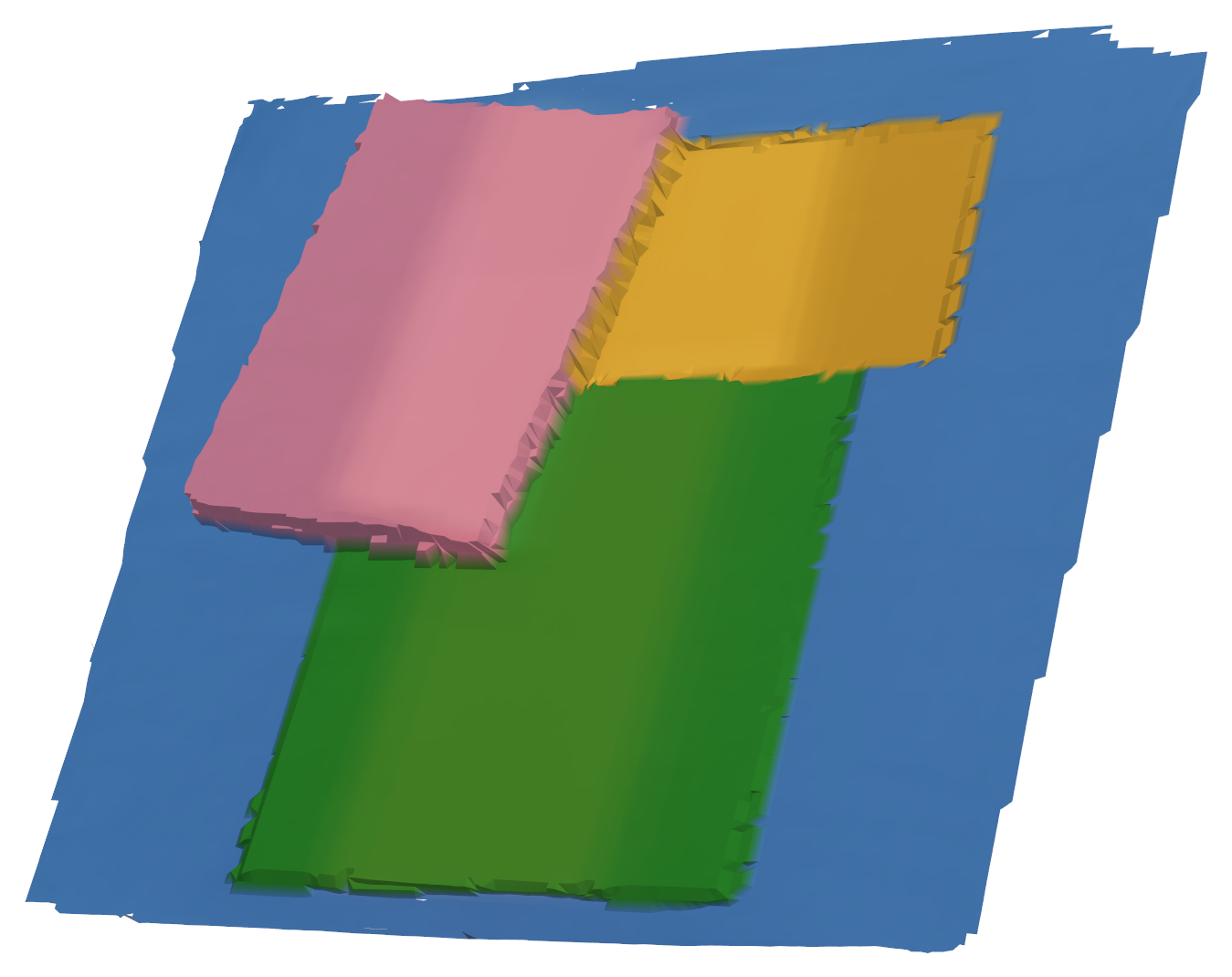}\\
  \end{tabular}
  \caption{The six captured study scenes: photograph of the physical
  scene (top), reconstruction (middle), and the automatic layer
  proposal (bottom; colors are layer identity, assigned in a fixed
  order). Per-scene VLM top-1 material accuracy against experimenter
  ground truth is annotated; the two before/after scenes (S1, S6) sit
  at 75\% and 50\%, giving editing headroom without a floor; the full
  pool spans 25--100\%.}
  \Description{A three-by-six grid of study scenes. The top row shows
  photographs of the physical tabletop scenes; the middle row shows
  reconstructed meshes of the same scenes, with household objects such
  as foam blocks, cardboard, paper, plastic and fabric on desks; the
  bottom row shows the same meshes with colored automatically proposed
  material regions. Column titles give each scene's role (before/after,
  freeform canvas, training, reserve) and VLM top-1 accuracy ranging
  from 25 to 100 percent.}
  \label{fig:scenes}
\end{figure*}

The study had three stages spanning replication and expression. In the
first two, participants compared an automatic draft with its physical scene,
rated its realism, edited until satisfied, and rated the result, intent
match, control, and workload. The two scenes were counterbalanced and chosen
from the six-scene benchmark (Figure~\ref{fig:scenes},
Table~\ref{tab:scenes}): exact-material accuracy was 75\% for the easier
scene and 50\% for the harder. Training used a separate scene.

In the third stage, participants began with an empty layer stack and used
the same operations without an automatic draft or physical referent. They
rated name--feel match, expressivity, and workload. This exploratory stage
examined expression rather than serving as a replication baseline.

An earlier pilot included scratch, automatic-only, and correction conditions
to compare time and workload. Participants treated scratch authoring as
creation rather than replication, making those comparisons uninterpretable.
We excluded the pilot sessions and instead compare automatic with revised
drafts within participant; freeform authoring remains a separate exploratory
task.

\subsection{Participants and Apparatus}
Twelve right-handed participants (7 men, 5 women; ages 18--34;
Table~\ref{tab:demographics}) took part under approval
from the Anonymous University Institutional Review Board. Participation was
voluntary and uncompensated, and all participants gave informed consent. Eight had little or no prior haptics experience,
and none were haptic-design experts; five had no 3D-tool experience. Sessions
lasted 40--80 minutes.

Participants used the browser interface and haptic display described in
\S\ref{sec:system} (Figure~\ref{fig:apparatus}). During replication, the
physical objects remained within reach for comparison. Scenes were
reconstructed beforehand, questionnaires ran in the browser, and TouchTwin
logged editing operations. Appendix~\ref{app:instruments} provides the
instruments.

\subsection{Measures}
For replication, the primary outcome was \texttt{scene\_match} (1--7;
``how closely does the virtual scene feel to the real objects''), measured
before and after editing. Participants also rated intent match and control.
Secondary measures included raw NASA-TLX~\cite{hart1988development},
SUS~\cite{brooke1996sus,bangor2009determining}, authoring-mode preference,
time, and logged semantic and spatial edits.

For freeform authoring, \texttt{named\_match} measured whether each region
felt like its user-provided name. Participants also rated expressivity,
control, and workload. These self-ratings assess intent--output consistency,
not physical realism.

Artifact-based measures from saved layers included face-weighted
material-family accuracy, GT-material rank among the 100 library materials,
and generated-model distance between a generated response and the measured
GT model. Together they distinguish category fidelity (the appropriate kind
of material) from specimen fidelity (a particular sample's response).

\subsection{Analysis}
Our confirmatory analysis averaged each participant's auto-to-revised realism
change across the two scenes and tested it against zero using a paired
$t$-test and Wilcoxon signed-rank test; we report within-participant effect
size $d_z$. Per-scene descriptives and analyses of the 24 participant--scene
pairs, presentation order, freeform authoring, and artifacts were
exploratory and uncorrected.

The NASA-TLX Performance item is reverse-anchored (0 = perfect, 100 =
failure); two participants appeared to reverse it. We recoded those responses
as $100-x$. Leaving them unchanged or omitting the item did not alter the
interpretation. Only the final participant saw clarified anchors.

\section{Results}
\label{sec:results}

\subsection{The Automatic Draft Provides a Useful Starting Point}
Before editing, participants rated automatic drafts $4.00/7$
($SD{=}1.38$) for realism but $5.4/7$ as useful starting points. Comments
identified specific mismatches, including a smooth-feeling black region and
cardboard labeled as wood. Participants could diagnose these errors through
touch and revise the descriptions rather than author responses from scratch.

\subsection{Brief Revision Improves Perceived Realism}
\label{sec:c2}
\begin{figure}
  \centering
  \includegraphics[width=0.85\columnwidth]{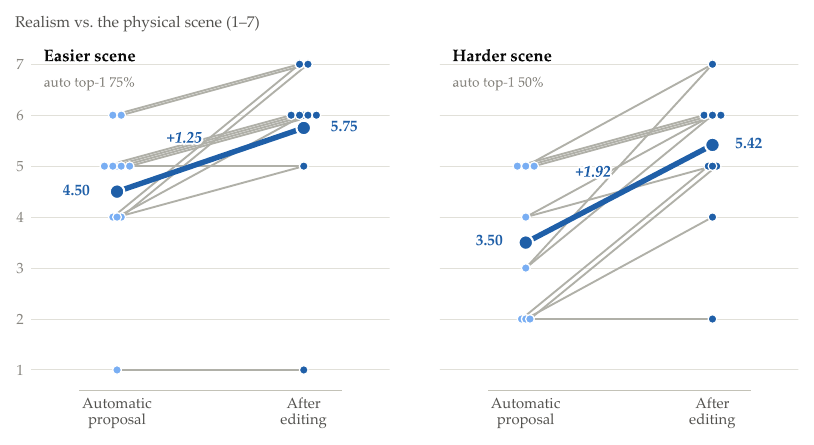}
  \caption{Realism versus the physical scene before (automatic proposal)
  and after participant correction, per scene ($N{=}12$). Thin lines are
  individual participants (overlapping pairs are dodged horizontally);
  the bold line is the mean with the paired lift annotated. No
  participant-scene pair was rated worse after editing.}
  \Description{Two slopegraphs, one per study scene. Each shows twelve
  participant lines rising from the automatic-proposal rating to the
  post-edit rating on a 1-to-7 realism scale. In the easier scene the
  mean rises from 4.50 to 5.75, a lift of 1.25; in the harder scene from
  3.50 to 5.42, a lift of 1.92. Light blue dots mark the automatic
  checkpoint and dark blue dots the post-edit checkpoint; a legend
  identifies both plus the bold mean line. No line decreases.}
  \label{fig:lift}
\end{figure}

Mean realism increased from $4.00/7$ before editing to $5.58/7$ afterward
(Figure~\ref{fig:lift}). The
mean participant-level change was $+1.58$ ($SD{=}0.73$),
$t(11){=}7.48$, $p{<}.0001$, $d_z{=}2.16$. Eleven participants improved,
one was unchanged, and none worsened (Wilcoxon on nonzero changes,
$p{=}.001$). Across 24 participant--scene pairs, 21 improved and none
worsened. In exploratory, uncorrected comparisons, the harder scene improved
by $+1.92$ points versus $+1.25$ for the easier; second scenes improved by
$+2.00$ versus $+1.17$ for first scenes, though scene and order are
confounded.

Revisions took a median of $6.6$ minutes per scene (IQR $5.3$--$9.6$).
Participants rated intent match $5.8/7$ and control $5.5/7$ after editing;
mean recoded raw NASA-TLX was $21.7/100$. Thus, brief revision produced a
large perceived-realism gain with relatively low workload.

\subsection{Participants Primarily Revise through Language}
In 14 of 24 replication trials, participants only relabeled materials; the
others combined relabeling with brush strokes, SAM-assisted expansion, or
face transfer; boundary repairs concentrated on the easier scene (8 of those
10 trials). Named layers therefore often turned an incorrect material
assignment into a semantic correction rather than a geometry edit.

The logs contained 294 generated-label events (247 relabels and 47 labels
for new layers). After removing partial prefixes recorded as typing settled,
150 committed descriptions remained; 69 labels appeared in final scenes.
Of those final labels, 91\% were multi-word and 91\% used vocabulary absent
from the 100 library names. They ranged from ``cardboard'' to ``loud, black
plastic sheet, super low friction, grainy like super fine sand.'' However,
we did not compare free text with a searchable picker, and open-vocabulary
use was not associated with realism improvement (Spearman $\rho{=}.16$,
$p{=}.62$). These data describe how participants revised drafts, not whether
free text caused the realism gain.

\begin{figure}
  \centering
  \includegraphics[width=0.95\columnwidth]{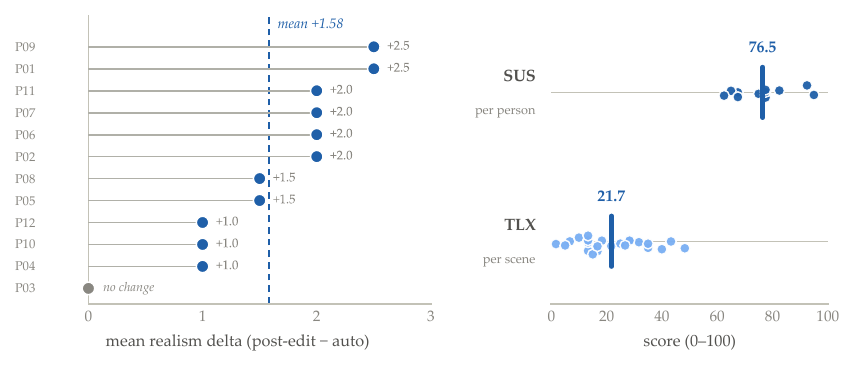}
  \caption{Left: mean realism delta per participant (dotted line =
  grand mean; P03, gray, is the one unchanged participant). Right: SUS
  per participant and raw NASA-TLX per scene-trial; diamonds are
  means.}
  \Description{Left panel: a lollipop chart of twelve participants
  sorted by their mean realism improvement, ranging from 0 for P03 to
  2.5 for P09 and P01, with the mean of 1.58 marked. Right panel: strip
  plots of SUS scores clustered between 62 and 95 with mean 76.5, and
  TLX workload scores clustered between 2 and 50 with mean 21.7.}
  \label{fig:participants}
\end{figure}

\subsection{Revision Improves Category but Not Specimen Fidelity}
\label{sec:objective}
\begin{figure}
  \centering
  \includegraphics[width=0.85\columnwidth]{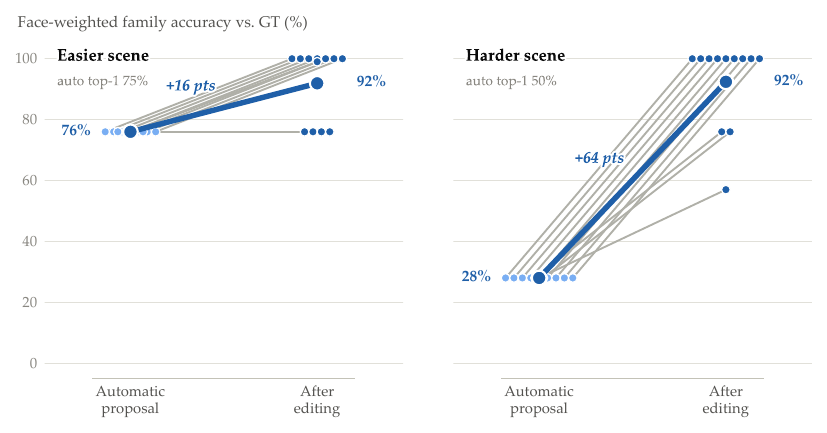}
  \caption{The artifact-based counterpart to Figure~\ref{fig:lift}:
  \emph{face-weighted} material-family accuracy of each participant's
  layer stack against experimenter ground truth, before and after
  editing. This score is computed from the saved artifacts and does not
  depend on participants' self-reports. Area-weighted, so not comparable
  to the per-tile Family column of Table~\ref{tab:what}. No trial worsened.}
  \Description{Two slopegraphs matching the layout of the realism
  figure, showing face-weighted material-family accuracy versus ground
  truth. In the easier scene the mean rises from 76 to 92 percent; in
  the harder scene from 28 to 92 percent. All individual lines rise or
  stay flat; none fall.}
  \label{fig:objective}
\end{figure}

Because participants knew they had edited before re-rating, we complemented
their realism ratings with measures computed from the saved material layers
(Figure~\ref{fig:objective}). Face-weighted material-family accuracy rose
from 52\% to 92\% (easier scene $76{\rightarrow}92\%$; harder
$28{\rightarrow}92\%$), improving in 20 of 24 trials and worsening in none
(Wilcoxon $p{<}.0001$). This area-weighted metric is not comparable to the
per-tile column in Table~\ref{tab:what}. The rank of the true name among
100 library materials improved in 22 trials (harder-scene median
$35{\rightarrow}4.7$), but this measure uses the same CLIP space as the
system and is not independent.

Editing did not consistently move the generated haptic responses closer to
the measured specimens. Generated-model distance changed from $0.22$ to
$0.18$ on the easier scene
(9/12 improved, $p{=}.33$) and from $0.52$ to $0.66$ on the harder one
(4/12, $p{=}.56$). For context, generating each of the 13 GT materials
from its exact library name produced a mean distance of $0.092$ (median
$0.051$, range $0.002$--$0.274$; random pairs: $0.834$). The generator can
therefore closely reproduce many measured library materials when given their
exact names. On the harder scene, however, editing increased distance from
the specific specimen even as realism and family accuracy improved.

Participants usually entered descriptions rather than library names, so a
label could identify the appropriate kind of material yet remain distant
from the recorded GT specimen. Revision therefore improved category fidelity
without establishing fidelity to the individual physical specimen.

\subsection{From Replication to Freeform Haptic Authoring}
\label{sec:c3}
\begin{figure}
  \centering
  \includegraphics[width=0.6\columnwidth]{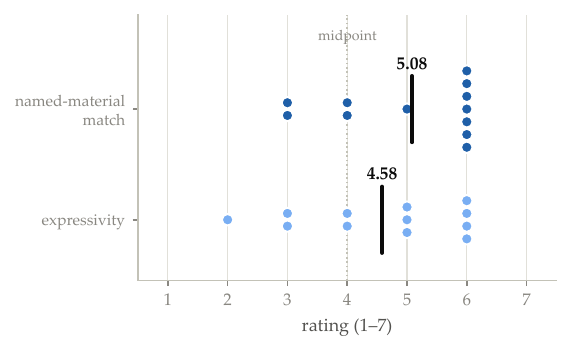}
  \caption{Freeform stage ($N{=}12$, no physical referent): per-region
  named-material match and overall expressivity; diamonds are means,
  dotted line the scale midpoint.}
  \Description{Strip plot of twelve named-material match ratings
  between 3 and 6 with mean 5.08, and twelve expressivity ratings
  between 2 and 6 with mean 4.58, on a 1 to 7 scale with the midpoint
  4 marked.}
  \label{fig:freeform}
\end{figure}

In the freeform stage, participants began with an empty layer stack and
created a median of four material regions in 3.8 minutes. They rated
named-material match at $5.08/7$ ($SD{=}1.24$), above the scale midpoint
($t(11){=}3.03$, $p{=}.012$); expressivity averaged $4.58/7$
(Figure~\ref{fig:freeform}). Because participants evaluated their own
descriptions without a physical referent, these ratings indicate perceived
intent--output consistency, not that another person would identify the
generated materials in the same way.

Two participants rated named-material match at $3/7$; one said the regions
felt too similar. The generated signals nevertheless remained distinct:
across 64 freeform-layer pairs, median signal distance was $0.62$, and only
6\% fell below the minimum distance among automatic proposals ($0.24$).
Some signal differences may therefore have been insufficiently salient
through the stylus.

Preferences depended on the authoring goal. Nine participants preferred
starting from scratch for creative authoring, two preferred correcting an
automatic proposal, and one was indifferent. Their comments favored drafts
when a physical scene supplied a replication target, but an empty scene when
they wanted expressive control.

\subsection{Usability and Participant Perspectives}
\label{sec:preference}
SUS averaged $76.5/100$ (``good''; range $62.5$--$95$), and freeform
NASA-TLX averaged $19.9/100$. Four themes in participants' comments further
characterized the experience. \textbf{(T1) Roughness}: virtual textures often
felt bumpier than their physical counterparts. \textbf{(T2) Near-miss
recognition}: errors such as cardboard labeled as wood were noticeable
through touch and correctable through relabeling. \textbf{(T3) Missing
compliance}: the rigid proxy could not reproduce bulk softness such as
that of foam. \textbf{(T4) Tool mediation}: a stylus conveyed fewer cues than
direct hand contact. Thus, semantic errors could often be revised, whereas
missing display cues could not be recovered by changing a description.

\section{Discussion}
\label{sec:discussion}

\subsection{Imperfect AI Can Be Useful When Its Output Is Experiential and Editable}
TouchTwin's central interaction finding is not that its automatic material
recognition was consistently correct, but that an imperfect draft remained
useful when users could experience and revise it. Touch exposed mismatches
that looked plausible on screen, and participants usually corrected them by
renaming a region rather than rebuilding its geometry or tuning a signal.
This makes the target sense part of the feedback loop: the generated output
is inspected through touch, and the resulting perceptual judgment guides the
next edit. The pattern extends mixed-initiative interaction
principles~\cite{horvitz1999principles,amershi2019guidelines} by suggesting
that sensory AI output should be both editable and directly experienceable.
For replication tasks, a person who can access the physical object could
revise a scan before sharing it with remote shoppers, museum visitors, or
collaborators. The supplemental video illustrates this workflow on stock
videos of an antique typewriter and a fabric scene, although these examples
are demonstrations rather than evaluations. Such uses fit TouchTwin's
category-level goal; inspection or quality-control applications would
additionally require specimen- and defect-level fidelity.

\subsection{Language as a Shared Representation for Human--AI Haptic Authoring}
Language did more than replace sliders. Material descriptions formed a
shared representation across the automatic proposal, the interface, and the
haptic generator. The system proposed a named region, the user interpreted
and edited that name, and the generator converted the revised description
into a response. Users therefore did not have to translate a felt mismatch
into spectral, friction, or force parameters. Their frequent use of
multi-word descriptions outside the library vocabulary also suggests that
the shared representation can accommodate distinctions users naturally want
to express. However, the study did not compare free text with a searchable
material picker, so it does not establish that language is always the most
efficient input method. Its demonstrated value here is continuity between
human and AI representations: both parties act on the same understandable
unit while a model grounded in measured interactions handles the mapping to
renderable signals~\cite{qian2026language}.

\subsection{Different Roles for AI in Replication and Expression}
Participants wanted different levels of AI initiative for different goals.
When reproducing a physical scene, the automatic proposal provided a
starting point and hypotheses that could be checked against a referent. When
creating a new haptic scene, most participants preferred an empty canvas,
where suggestions were less useful than control over what to express. This
distinction argues against a single default role for generative assistance:
AI can lead with a draft when a target exists, but recede to optional
suggestions when intent is still being formed. The latter turns surface-level
haptic authoring into a descriptive task---for example, allowing a VR author
to explore ``brushed aluminum'' and ``powder-coated steel'' without first
learning signal design~\cite{schneider2017haptic,seifi2019haptipedia}. More
generally, sensory authoring tools should adapt initiative to whether users
are matching an external referent or developing an expressive one.

\subsection{Design Implications for Human--AI Sensory Authoring}
\textbf{Align the AI output with the user's correction unit.} In TouchTwin,
a named material region is simultaneously what the system infers, what the
user feels, and what the user can revise. This alignment turns an inference
error into a local correction instead of forcing users to discard the draft
or manipulate a lower-level representation. Other sensory authoring systems
could apply the same principle by exposing sound sources, lighting regions,
or physical behaviors as editable units whose boundaries and meanings remain
legible throughout the pipeline.

\textbf{Ground open-ended controls in measured experience.} Language gives
users a broad vocabulary, but a sensory interface must ultimately produce a
physical signal. TouchTwin maps descriptions through a model trained on
measured tool--surface interactions, allowing expressive input while keeping
output tied to the dynamics the renderer can represent. This grounding does
not guarantee a match to a particular specimen, and unsupported properties
such as compliance cannot be recovered by more descriptive wording. Designers
should therefore make the coverage of the underlying measurements visible and
combine open-ended input with bounded alternatives when predictable output is
more important than expressivity.

\textbf{Evaluate both editable outcomes and fidelity beyond the interaction
loop.} Participants could directly optimize whether a scene felt plausible
and whether regions represented the intended material category; both
measures improved after editing. Distance to the recorded physical specimen
did not reliably improve, because users neither saw that metric nor entered exact
library identities. Evaluations of sensory AI should likewise separate
perceived quality, task-level semantic agreement, and physical signal
fidelity. Reporting these levels together clarifies what human revision can
repair, what remains a renderer or sensing limitation, and which application
claims the evidence supports.

\section{Limitations and Future Work}
\label{sec:limitations}

\textbf{Participants, scenes, and uses.} Our 12 participants were ages
18--34, mostly engineering students, at one site with one device. The two
before/after scenes are fixed effects and do not support generalization
across scenes. All six captures are near-planar tabletop arrangements,
favorable to reconstruction, mask lifting, and VLM cropping; curvature,
clutter, occlusion, specularity, and thin structures remain untested.
Participants also saw visibly blurry reconstructions while rating feel,
creating a possible visual confound. Future studies should include more
diverse participants, devices, scenes, and capture conditions. Touchable
scans may eventually support nonvisual access, but our participants were
sighted and a stylus may not be the preferred interface; that direction
requires participatory design with blind and low-vision users.

\textbf{Measures, validity, and ground truth.} Participants knew they had
edited before re-rating each scene, so demand effects could inflate the
subjective improvement. The physical objects anchored both ratings, three
participant--scene pairs remained unchanged, and material-family accuracy
computed from the saved layers increased from 52\% to 92\%. This convergence
strengthens the realism result but does not eliminate the possible bias.
Neither measure establishes fidelity to a particular specimen: a description
can identify the right category without producing the signal nearest to one
recorded sample. We also defined the ten material families, and another
taxonomy could change the family-level results. Some discontinued library
products were replaced by another brand of the same category, so the measured
GT exemplar did not always match the object on the desk. Editing did not
improve generated-model distance on the harder scene
(\S\ref{sec:objective}), and realism was measured per scene rather than per
region. A rated oracle-label condition would separate recognition error from
the remaining rendering gap, while a searchable library-picker baseline
would test the contribution of free-text input. Per-region psychophysics,
physical-sample 2AFC comparisons, and blind cross-participant identification
would provide stronger tests of name--feel correspondence.

\textbf{Recognition and rendering scope.} The automatic proposals selected
the exact material for only 13 of 24 surfaces in our small benchmark, so
fidelity-sensitive uses still require human review and revision. The model
also cannot represent bulk compliance in food or soft tissue, the hand of a
fabric, or thermal cues. Holding each 1\,kHz texture sample across the audio
buffer without reconstruction filtering introduces energy in the Pacinian
band and may explain reports of excess roughness
(\S\ref{sec:rendering}). Layer transitions do not cross-fade, and we did
not validate our use of the contact binormal rather than the surface normal.
Two participants found distinct textures hard to distinguish through the
stylus, and reconstruction errors also remain consequential. Follow-up work
should measure coil spectra and end-to-end latency, add transition filtering,
validate rendering direction, and incorporate compliance or thermal displays
where the application requires those cues.

\textbf{Authoring at scale.} A median 6.6-minute edit is plausible for a
single product or small scene but costly across a room or building, and the
present workflow assumes that one person revises one layer stack. Larger
deployments will require faster on-device processing, reusable or shared
material layers, and higher-level operations such as ``make everything on
the left metallic.'' Evaluating those mechanisms on larger scenes and in
multi-author workflows is necessary before the approach can support
environment-scale authoring.

\section{Conclusion}
TouchTwin treats AI-generated haptic content as an experiential, editable
draft rather than a final prediction. It turns a phone scan into named
material regions that people can feel and revise using the same semantic
representation as the automation. In our study, participants used this
propose--feel--revise loop to improve imperfect replicas and used the same
mechanism to create new material experiences, while preferring different
levels of AI initiative for replication and expression. These findings
suggest a broader direction for human--AI sensory authoring: align the AI's
output with the user's correction unit, ground open-ended input in measured
experience, and evaluate both outcomes users can perceive and fidelity
limits they cannot.


\bibliographystyle{ACM-Reference-Format}
\bibliography{references}

\appendix
\makeatletter
\@addtoreset{figure}{section}
\@addtoreset{table}{section}
\makeatother
\renewcommand{\thefigure}{\thesection.\arabic{figure}}
\renewcommand{\thetable}{\thesection.\arabic{table}}
\renewcommand{\theHfigure}{appendix.\thesection.\arabic{figure}}
\renewcommand{\theHtable}{appendix.\thesection.\arabic{table}}
\section{Supplementary Material}
\subsection{Study Instruments}
\label{app:instruments}
All questionnaires were administered in-browser; items below are reproduced verbatim from the study software (the NASA-TLX Performance item shows the clarified reverse-anchor wording used from P12 onward; earlier participants saw ``perfect/failure'' anchors only).

\subsubsection*{Before editing --- the automatic result}
\emph{You just felt the scene the SYSTEM proposed automatically, with no edits. Compare what you felt to the real objects in front of you.}

\begin{itemize}
\item How closely does the scene you just FELT match the REAL objects in front of you? \\ {\small [1--7 (not at all $\to$ matches exactly)]}
\item Which regions feel WRONG vs the real objects (if any)? (optional) \\ {\small [free text]}
\end{itemize}

\subsubsection*{After editing --- your final result}
\emph{Now rate the scene you FELT after your edits, against the real objects.}

\begin{itemize}
\item How closely does the scene you just FELT match the REAL objects in front of you? \\ {\small [1--7 (not at all $\to$ matches exactly)]}
\item The finished scene feels the way I intended it to. \\ {\small [1--7 (strongly disagree $\to$ strongly agree)]}
\item I was able to get the result I wanted with the editing tools. \\ {\small [1--7 (strongly disagree $\to$ strongly agree)]}
\item NASA-TLX --- Mental demand \\ {\small [0--100 slider (very low $\to$ very high)]}
\item NASA-TLX --- Physical demand \\ {\small [0--100 slider (very low $\to$ very high)]}
\item NASA-TLX --- Temporal (time pressure) \\ {\small [0--100 slider (very low $\to$ very high)]}
\item NASA-TLX --- Performance: how successful were you? NOTE --- this scale is reversed: 0 = perfect, 100 = failure (LOWER is better). \\ {\small [0--100 slider (0 = perfect $\to$ 100 = failure)]}
\item NASA-TLX --- Effort \\ {\small [0--100 slider (very low $\to$ very high)]}
\item NASA-TLX --- Frustration \\ {\small [0--100 slider (very low $\to$ very high)]}
\item Any quick comments on this scene (optional): (optional) \\ {\small [free text]}
\end{itemize}

\subsubsection*{After free authoring}
\emph{You authored textures freely by naming materials (no real scene to match). Rate the felt result against what each NAME should feel like.}

\begin{itemize}
\item Each region felt like the material I NAMED it. \\ {\small [1--7 (not at all $\to$ very much)]}
\item I could create the textures I imagined. \\ {\small [1--7 (strongly disagree $\to$ strongly agree)]}
\item Authoring patches from scratch felt controllable. \\ {\small [1--7 (strongly disagree $\to$ strongly agree)]}
\item NASA-TLX --- Mental demand \\ {\small [0--100 slider (very low $\to$ very high)]}
\item NASA-TLX --- Physical demand \\ {\small [0--100 slider (very low $\to$ very high)]}
\item NASA-TLX --- Temporal (time pressure) \\ {\small [0--100 slider (very low $\to$ very high)]}
\item NASA-TLX --- Performance: how successful were you? NOTE --- this scale is reversed: 0 = perfect, 100 = failure (LOWER is better). \\ {\small [0--100 slider (0 = perfect $\to$ 100 = failure)]}
\item NASA-TLX --- Effort \\ {\small [0--100 slider (very low $\to$ very high)]}
\item NASA-TLX --- Frustration \\ {\small [0--100 slider (very low $\to$ very high)]}
\item Any comments on free authoring (optional): \\ {\small [free text]}
\end{itemize}

\subsubsection*{System Usability Scale (full system)}
\emph{Thinking about the FULL tool (auto-inferred layers + editing + relabel). 1 = strongly disagree, 5 = strongly agree.}

\begin{itemize}
\item I think that I would like to use this system frequently. \\ {\small [1--5 (strongly disagree $\to$ strongly agree)]}
\item I found the system unnecessarily complex. \\ {\small [1--5 (strongly disagree $\to$ strongly agree)]}
\item I thought the system was easy to use. \\ {\small [1--5 (strongly disagree $\to$ strongly agree)]}
\item I think that I would need the support of a technical person to use this system. \\ {\small [1--5 (strongly disagree $\to$ strongly agree)]}
\item I found the various functions in this system were well integrated. \\ {\small [1--5 (strongly disagree $\to$ strongly agree)]}
\item I thought there was too much inconsistency in this system. \\ {\small [1--5 (strongly disagree $\to$ strongly agree)]}
\item I would imagine that most people would learn to use this system very quickly. \\ {\small [1--5 (strongly disagree $\to$ strongly agree)]}
\item I found the system very cumbersome to use. \\ {\small [1--5 (strongly disagree $\to$ strongly agree)]}
\item I felt very confident using the system. \\ {\small [1--5 (strongly disagree $\to$ strongly agree)]}
\item I needed to learn a lot of things before I could get going with this system. \\ {\small [1--5 (strongly disagree $\to$ strongly agree)]}
\end{itemize}

\subsubsection*{Wrap-up}
\emph{You corrected the system's automatic proposal on two scenes, and authored textures freely on a third.}

\begin{itemize}
\item The automatic proposal was a useful starting point to edit from. \\ {\small [1--7 (strongly disagree $\to$ strongly agree)]}
\item Which did you find more satisfying? \\ {\small [one of: correcting the automatic proposal; authoring freely from scratch; about the same]}
\item Briefly, why? \\ {\small [free text]}
\end{itemize}

\subsubsection*{Semi-structured interview}
\emph{Open questions --- capture the participant's answers (paraphrase ok).}

\begin{itemize}
\item Where did the AUTOMATIC result help you most? \\ {\small [free text]}
\item Where did the automatic result fail or surprise you? \\ {\small [free text]}
\item How did selecting/correcting REGIONS feel? Smooth or frustrating, and where? \\ {\small [free text]}
\item How did changing a material by TYPING a name feel? \\ {\small [free text]}
\item Did the rendered textures match the real objects? Where not? \\ {\small [free text]}
\item Did you trust the felt output? Why / why not? \\ {\small [free text]}
\item What would you change or add? \\ {\small [free text]}
\item Anything else? (optional) \\ {\small [free text]}
\end{itemize}

\subsection{Scene Details}
\label{app:scenes}
Six tabletop scenes were captured with the same ${\sim}30$\,s phone-orbit protocol (Figure~\ref{fig:scenes}). Physical objects span foam, wood, cardboard, paper, plastic, fabric, and leather on an oak desk.

\begin{table}[h]
\caption{Study scenes: role, mesh size, number of auto-proposed layers, and the VLM's top-1 material accuracy against experimenter ground truth.}
\label{tab:scenes}
\begin{tabular}{llrrrr}
\toprule
Scene & Role & Vertices & Faces & Layers & VLM top-1\\
\midrule
S1 & before/after (easier) & 21,515 & 42,146 & 4 & 75\%\\
S2 & reserve & 20,556 & 40,455 & 4 & 25\%\\
S3 & reserve & 27,191 & 53,043 & 4 & 25\%\\
S4 & freeform canvas & 20,054 & 39,466 & 4 & 50\%\\
S5 & training & 16,094 & 31,509 & 4 & 100\%\\
S6 & before/after (harder) & 17,183 & 33,699 & 4 & 50\%\\
\bottomrule
\end{tabular}
\end{table}

\subsection{Participant Demographics}
\label{app:demographics}
\begin{table}[h]
\caption{Participant demographics ($N{=}12$, all right-handed, all screened for normal tactile sensitivity).}
\label{tab:demographics}
\begin{tabular}{ll}
\toprule
Age & 18--24: 3;\quad 25--34: 9\\
Gender & 7 men, 5 women\\
Haptics experience & none: 4;\ tried once or twice: 4;\ occasional: 2;\ regular: 2\\
3D-tool experience & none: 5;\ beginner: 3;\ intermediate: 4\\
Background (multi) & engineering: 7;\ student: 8;\ haptics: 2;\ design/HCI/graphics: 3;\ other: 2\\
\bottomrule
\end{tabular}
\end{table}

\subsection{System Parameters and Reproducibility}
\label{app:params}

All values below are the ones used for every study session.

\begin{table}[h]
\caption{Models and pipeline parameters.}
\label{tab:params}
\begin{tabular}{ll}
\toprule
\multicolumn{2}{l}{\emph{Models (all frozen; no component was fine-tuned)}}\\
Geometry & VGGT (feed-forward multi-view transformer)\\
Depth completion & SfM-guided monocular depth diffusion + TSDF\\
Segmentation & SAM~2.1 Hiera-Tiny, prompt + automatic mask modes\\
Material naming & Qwen2.5-VL-3B-Instruct, greedy, $\le$40 new tokens\\
Text embedding & open\_clip ViT-L/14, \texttt{openai} weights (768-D)\\
Text$\rightarrow$haptic & language-aligned generative decoder~\cite{qian2026language}\\
\midrule
\multicolumn{2}{l}{\emph{Generated haptic model (per layer, as stored)}}\\
AR grid & 18 (speed, force) nodes\\
AR representation & 20 line-spectral frequencies + excitation variance\\
Tap set & 13 speed-binned envelopes, 100 samples each\\
\midrule
\multicolumn{2}{l}{\emph{Renderer (1\,kHz servo)}}\\
Proxy stiffness / damping & 1.0 / 0.35 (normalized device units)\\
Contact spring & 220\,N per scene unit, clamped at 4\,N\\
Pop-through & 0.9 of max device force; depth 0.28\\
Texture gain / ceiling & 0.05 on raw AR; 3.0\,N soft ceiling\\
Tap gain & 0.025\\
Friction & measured $\mu$, viscous slope 0.004 per mm/s\\
Slide threshold & 5\,mm/s (below this, no texture is emitted)\\
Contact threshold & 0.12\,N\\
Velocity filter & Butterworth, $\lambda = 125$\,rad/s\\
Workspace scale & device workspace spans 0.25 of scene bbox\\
Audio actuator & MM3C Haptuator (TactileLabs)\\
Audio channel & WASAPI 48\,kHz, $\sim$1\,ms buffers, gain 0.02, texture only\\
\bottomrule
\end{tabular}
\end{table}

\textbf{VLM prompt (verbatim).} ``What is the surface MATERIAL and
TEXTURE of the main object in this image? Answer with 3 to 6 words
describing the material and how it would feel to touch (for example:
`smooth varnished oak wood', `brushed silver metal', `soft black foam',
`fine grained brown leather'). Describe the material, not the object's
name.'' Captions failing a material-vocabulary filter are discarded and
the region is left unlabeled rather than guessed. Table~\ref{tab:vlmio}
gives every VLM output on the two before/after scenes verbatim, paired
with the library material it resolved to and the ground truth, per
recent guidance on reporting LLM-integrated
systems~\cite{navarro2026reporting}: the three errors are visibly
near-misses rather than random guesses---rigid ABS plastic read as
stained wood, a velvet surface read as plain wood, and (again) soft
EPDM foam read as rigid ABS plastic, the same soft-reads-as-hard
pattern reported in \S\ref{sec:eval-what}---which is the basis for
that section's claim that correction is cheap.

\begin{table}[h]
\caption{Every automatic proposal on the two before/after scenes: the
VLM's verbatim caption, the library material it resolved to, and
ground truth. Boldface marks wrong labels: 1 of 4 on S1 (75\% top-1)
and 2 of 4 on S6 (50\%), matching Table~\ref{tab:what}.}
\label{tab:vlmio}
\begin{tabular}{p{0.44\columnwidth}ll}
\toprule
VLM caption (verbatim) & Resolved & GT\\
\midrule
\multicolumn{3}{l}{\emph{S1 (candidate\_01)}}\\
smooth white foam & Styrofoam & Styrofoam\\
smooth varnished oak wood & Wood & Wood\\
smooth varnished dark wood & \textbf{Stained Wood} & \textbf{ABS Plastic}\\
smooth varnished oak wood & Wood & Wood\\
\midrule
\multicolumn{3}{l}{\emph{S6 (candidate\_06)}}\\
smooth varnished oak wood & \textbf{Wood} & \textbf{Velvet}\\
orange matte plastic & \textbf{ABS Plastic} & \textbf{EPDM Foam}\\
wooden clipboard, rough texture, matte finish & Cardboard & Cardboard\\
lined yellow paper, matte finish, soft texture & Resume Paper & Resume Paper\\
\bottomrule
\end{tabular}
\end{table}

\end{document}